\pdfoutput=1
\documentclass[journal]{IEEEtran}

\usepackage[T1]{fontenc}
\usepackage{textcomp}
\usepackage{amsmath,amsfonts,amssymb,mathtools}
\usepackage{cite}
\usepackage{booktabs,tabularx,array}
\usepackage{xcolor}
\usepackage{graphicx}
\usepackage{dblfloatfix}
\usepackage{algorithm}
\usepackage{algpseudocode}
\usepackage{microtype}
\usepackage{xurl}
\usepackage[hidelinks,bookmarks=true]{hyperref}
\hypersetup{
  pdftitle={VeriSegue: Local RL and Cloud LLM Collaborative Verification Across RISC-V Processors and AI Accelerators},
  pdfauthor={Yiran He, Haihua Shen, Pei Chen, Shan Li, and Huawei Li},
  pdfkeywords={Hardware functional verification, coverage-directed test generation, RISC-V processors, AI accelerators, large language models, reinforcement learning, differential checking}
}
\newcommand{\nangate}{Nangate\allowbreak Open\allowbreak Cell\allowbreak Library}
\newcommand{\axadder}{\texttt{adder\_\allowbreak fpany\_\allowbreak no\_\allowbreak norm\_\allowbreak v2.v}}
\newcommand{\tcode}[1]{\text{\scriptsize\ttfamily #1}}

\newcommand{\circnum}[1]{\textcircled{\raisebox{-0.7pt}{\scriptsize\textbf{#1}}}}

\let\oldthebibliography\thebibliography
\renewcommand\thebibliography[1]{\oldthebibliography{#1}\setlength{\itemsep}{-1.1pt}\setlength{\parskip}{0pt}}

\begin{document}
\bstctlcite{bstctl:tcad}

\title{VeriSegue: Local RL and Cloud LLM Collaborative Verification Across RISC-V Processors and AI Accelerators}

\author{Yiran~He, Haihua~Shen, Pei~Chen, Shan~Li, and~Huawei~Li%
\thanks{\textit{(Corresponding author: Haihua Shen.)}}%
\thanks{Yiran~He, Haihua~Shen, and Pei~Chen are with the School of Computer Science and Technology, University of Chinese Academy of Sciences, Beijing 100049, China (e-mail: heyiran231@mails.ucas.ac.cn; shenhh@ucas.ac.cn; chenpei24@mails.ucas.ac.cn).}%
\thanks{Shan~Li is with the College of Intelligence Science and Technology (College of Cyberspace Security), Inner Mongolia University of Technology, Hohhot 010080, Inner Mongolia, China (e-mail: lishan203@mails.ucas.ac.cn).}%
\thanks{Huawei~Li is with the State Key Laboratory of Processors, Institute of Computing Technology, Chinese Academy of Sciences, Beijing 100190, China (e-mail: lihuawei@ict.ac.cn).}}

\markboth{PREPRINT}%
{HE \MakeLowercase{\textit{et al.}}: VERISEGUE: LOCAL RL AND CLOUD LLM COLLABORATIVE VERIFICATION}

\maketitle

\begin{abstract}
Pre-silicon verification of RISC-V processors and AI accelerators faces a trade-off between local throughput and deep-state reachability: local reinforcement-learning (RL) generators stall before multi-step preconditions, whereas one-way large language model (LLM) pipelines suffer from template ceilings, retry latency, hidden offsets, parameter collapse, and cold-reset restarting. We present VeriSegue, a closed-loop local--cloud framework pairing compact on-premise RL with execution-guided frozen-LLM reasoning. Grounded in our findings, this paradigm points to a promising direction across chip design and verification, keeping proprietary RTL and waveforms local for R\&D privacy while uniting local speed with bounded cloud reasoning for stronger effectiveness. VeriSegue triggers handoff via control-flow velocity ($\gamma_{\mathrm{cf}}$) and returns deep contexts to local rollouts at zero extra cloud cost; Stage~2 integrates backward-slice CF-UCB scheduling, constructor synthesis (\texttt{SynLib}), hybrid formal/trace inversion (TraceSMT), and boundary batching under a three-oracle gate. Across five open-source RTL designs (RocketChip, SonicBOOM, Gemmini, AxCore, and unseen CVA6), eight LLMs, and Verilator/VCS, Stage-1 GRPO outperforms masked GenHuzz PPO on all five VCS metrics in $1.96\times$ less time. Under equal wall-clock horizons, VeriSegue wins $5/5$ seeds ($p=0.0079$) over FLAME on every target, lifting real RTL coverage by $+20.2\%$--$41.4\%$ ($+9.17\,$pp VCS Score), reaching $100\%$ zero-shot CVA6 functional closure, and cutting accelerator cost by $4.1\times$--$4.5\times$ ($9.0/10$ AxCore defects). Differential triage exposes $20$ new root causes ($11$ RTL bugs and live 0-day \texttt{GEM-D8}), $53$ silent Gemmini escapes, and a formally checked $45\,\mathrm{nm}$ AxCore patch eliminating a $-47.8\,$pp ImageNet-1K ResNet-18 collapse and $100\%$ quantized-LLM NaN failure.
\end{abstract}


\section{Introduction}
\label{sec:intro}

Pre-silicon functional verification dominates design effort across open-source RISC-V processors and AI accelerators, spanning scalar and out-of-order (OoO) cores, Rocket Custom Coprocessor (RoCC) systolic arrays, and approximate block-floating-point (block-FP) matrix engines~\cite{fine,cdgreview,rocketchip,sonicboom,gemmini,axcore}. Although these targets expose distinct stimulus interfaces (bare-metal instruction streams, C+RoCC coprocessor schedules, and block-FP matrix tiles), verification engineers encounter the same structural bottleneck across all of them. Fast local generators sweep shallow decode, pipeline, and datapath logic rapidly, yet residual coverage holes and severe silicon defects hide behind multi-step architectural preconditions, undocumented internal arithmetic offsets, and coupled boundary conditions. Exercising a RISC-V page-table walker (PTW) requires coherent multi-level virtual-memory tables and supervisor trap delegation. Exposing Gemmini's 20-bit output-stationary (OS) accumulator saturation or \texttt{LoopMatmul} underflow requires coordinated tile and schedule parameters, and triggering signed-minimum overflow (\texttt{INT\_MIN} $= -65{,}536$) in AxCore's normalizer requires inverting an undocumented internal alignment bias.

Combining the throughput of local simulation with the semantic depth required to enter deep hardware states remains difficult across heterogeneous designs. Existing automated verification paradigms hit two complementary ceilings.

\textbf{Ceiling~1: Cold-Reset Saturation in Local Fuzzers and Local RL.} Graybox RTL fuzzers~\cite{rfuzz,directfuzz,difuzzrtl,thehuzz,hypfuzz,profuzz,cascade} and on-premise learned generators~\cite{genhuzz,deepverifier,curefuzz,hifuzz,covr} co-execute at zero cloud latency, yet stall once shallow decode and arithmetic paths are covered. Per-test novelty drops by more than $100\times$, destabilizing Proximal Policy Optimization (PPO) value critics (yielding negative late explained variance, EV, even on native VCS) while leaving virtual-memory, trap-delegation, and corner arithmetic states unreachable. Moreover, existing local-RL and LLM verifiers such as GenHuzz~\cite{genhuzz} and FLAME~\cite{flame} are tailored to RISC-V processors and lack native support for heterogeneous AI accelerators.

\begin{figure*}[t]
  \centering
  \includegraphics[width=\textwidth]{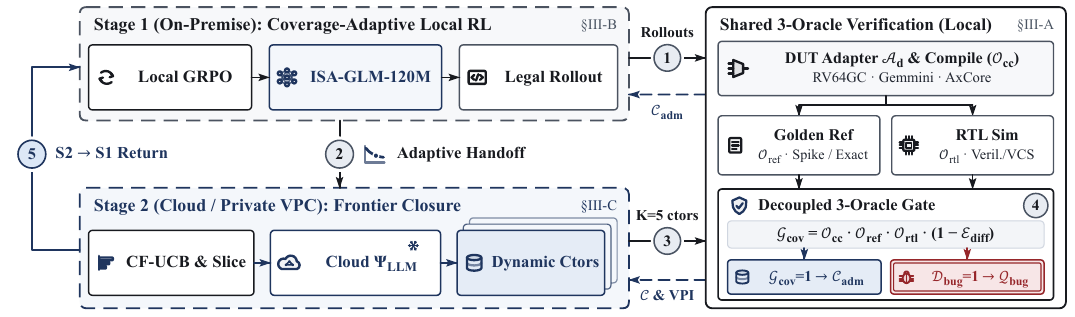}
  \caption{Closed-loop local--cloud verification architecture of VeriSegue. On-premise Stage~1 (\S\ref{sec:method_collab}, \circnum{1}) runs masked local GRPO until control-flow velocity $\gamma_{\mathrm{cf}}$ plateaus at $\tau^*$, handing off $\mathcal{C}_{\mathrm{adm}}^{(\tau^*)}$ (\circnum{2}) to frozen Stage~2 (\S\ref{sec:method_stage2}, \circnum{3}: CF-UCB, 4-hop slicing, TraceSMT, $K{=}5$ batching, and \texttt{SynLib}). Shared local verification (\S\ref{sec:method_infra}, \circnum{4}) decouples admitted coverage ($\mathcal{C}_{\mathrm{adm}}$) from bug triage ($\mathcal{Q}_{\mathrm{bug}}$, double-bordered crimson) while returning deep contexts $\mathcal{S}_{\mathrm{seed}}$ to Stage~1 at $\$0$ cloud cost (\circnum{5}). Solid/dashed arrows denote stimuli/feedback; snowflakes mark frozen weights.}
  \label{fig:architecture}
\end{figure*}

\textbf{Ceiling~2: Overhead and Reset Waste in One-Way LLM Pipelines.} While FLAME~\cite{flame} demonstrates the benefit of invoking LLMs with retrieval-augmented generation (RAG) after random processor exploration, one-way local-to-LLM hole closure suffers from four structural bottlenecks: (i)~\emph{retry-heavy open-loop prompting} that incurs low admission rates ($40.1\%$ on RocketChip) and static template ceilings; (ii)~\emph{blindness to hidden internal state} on deep sequential pipelines and biased datapaths; (iii)~\emph{greedy parameter collapse} onto nominal defaults (\texttt{rm=4}, \texttt{dim\_J=1}); and (iv)~\emph{one-way cold-reset waste}, which discards Stage-2 architectural contexts instead of seeding fast local rollouts.

To overcome both ceilings, we present VeriSegue (Fig.~\ref{fig:architecture}), a cross-architecture RTL verification framework that coordinates compact on-premise RL with execution-guided frozen-LLM reasoning through a shared three-oracle gate. Informed by our empirical findings, we further argue that such closed-loop local--cloud collaboration holds promising potential across broader chip design and verification tasks: retaining proprietary RTL and full waveforms on premise safeguards R\&D privacy, high-throughput local exploration accelerates convergence, and bounded cloud reasoning yields stronger verification effectiveness. This paper makes three core contributions:

\begin{enumerate}
  \item \textbf{Coverage-Adaptive Bidirectional Local--Cloud Collaboration:} A closed-loop verification formulation that couples critic-free, grammar-masked local RL (Stage~1) with frozen-LLM frontier closure (Stage~2). Rather than relying on hand-tuned simulation splits or cold-reset restarts, VeriSegue detects local saturation organically via toggle-immune control-flow velocity ($\gamma_{\mathrm{cf}}$) and returns verified deep architectural initialization contexts ($\mathcal{S}_{\mathrm{seed}}$) back to local rollouts at zero extra cloud queries.
  \item \textbf{Execution-Guided Frontier Closure for Deep Hardware States:} A Stage-2 closure engine integrating admission-decoupled, stall-bounded CF-UCB scheduling over compact backward RTL dossiers, sandboxed dynamic constructor synthesis (\texttt{SynLib}), hybrid formal pruning with near-miss trace inversion (TraceSMT), and boundary-diversified parameter batching, overcoming the static template ceilings, hidden internal offsets, and retry overhead of one-way LLM verification pipelines.
  \item \textbf{Unified CPU--Accelerator Architecture, Differential Triage, and Silicon Patching:} Unlike prior CPU-only local-RL and LLM verifiers (GenHuzz, FLAME), modular per-DUT adapters and a strict three-oracle gate separating coverage admission from bug triage enable one unified workflow across scalar/OoO RISC-V processors (including unseen SystemVerilog CVA6) and heterogeneous AI accelerators (Gemmini and AxCore) under both Verilator and VCS. Across all five targets, VeriSegue delivers consistent coverage and efficiency advantages while exposing $20$ newly discovered root causes, $53$ silent reference-model escapes, and a formally checked $45\,\mathrm{nm}$ AxCore patch that eliminates severe vision and quantized-LLM accuracy collapse.
\end{enumerate}

Section~\ref{sec:related} reviews related work, Section~\ref{sec:method} presents VeriSegue's methodology, Sections~\ref{sec:main_eval}--\ref{sec:generalization} report dual-simulator evaluation and generalization, Section~\ref{sec:defects_ai} audits discovered defects and AI workload impact, and Sections~\ref{sec:discussion}--\ref{sec:conclusion} discuss deployment and conclude.

\section{Background and Related Work}
\label{sec:related}

Automated RTL verification spans four main directions: graybox RTL fuzzing, architectural generation with golden-reference screening, local RL generators, and LLM-guided verification. Below we position VeriSegue against each family and identify the specific mechanisms required to unlock deep hardware states without excessive cloud retry cost.

\textbf{Graybox RTL Fuzzers and Formal-Assisted Seed Injection.} Moving beyond manual coverage models~\cite{fine,cdgreview}, graybox fuzzers steer simulation via structural feedback~\cite{verilator}. RFUZZ~\cite{rfuzz} tracks multiplexer toggles, DirectFuzz~\cite{directfuzz} weights module distances, DIFUZZRTL~\cite{difuzzrtl} monitors register finite-state machines (FSMs), and TheHuzz~\cite{thehuzz} adapts instruction mutations. Recent systems add multi-armed bandits (MABFuzz~\cite{mabfuzz}, ReFuzz~\cite{refuzz}), bounded model checking (BMC) or automatic test pattern generation (ATPG) seed injection (HyPFuzz~\cite{hypfuzz}, PROFUZZ~\cite{profuzz}), and timing, transient, or FPGA security fuzzing~\cite{whisperfuzz,fuzzitizer,dejavuzz_asplos2025,turbofuzz_hpca2026}. Rather than unrolling sequential BMC from reset over 100k-line cores, VeriSegue restricts Yosys cone-of-influence (COI) satisfiability (SAT) to fast local dead-clause pruning and shallow arithmetic solving, and reserves closest-cycle trace inversion (TraceSMT) for multi-cycle processor branches where bit-level SAT fails.

\textbf{Entangled Generation and Golden-Reference Screening.} Cascade~\cite{cascade} synthesizes open-loop entangled RISC-V programs without reading residual RTL holes, while GoldenFuzz~\cite{goldenfuzz} (aligned with recent oracle taxonomies~\cite{sok}) pre-screens instruction blocks on Spike~\cite{spike}. Fast instruction-set architecture (ISA) models neither expose uncovered RTL predicates nor detect their own specification omissions: as we show on Gemmini, \texttt{spike-gemmini} misses $53$ hardware assertions. VeriSegue embeds reference checks inside a strict three-oracle gate alongside an exact rational oracle for approximate arithmetic, keeping divergent runs isolated from coverage admission while preserving every bug witness.

\textbf{Local Learned Generators and Hierarchical RL.} ChatFuzz~\cite{chatfuzz}, GenHuzz~\cite{genhuzz} (GPT-2 PPO on VCS), DeepVerifier~\cite{deepverifier}, CURE-Fuzz~\cite{curefuzz}, CovR~\cite{covr} (reasoning-guided RL), and HiFuzz~\cite{hifuzz} (hierarchical semantic RL) provide fast policy learning from local simulation feedback at zero cloud latency. However, these generators focus exclusively on processor instruction streams without native support for heterogeneous AI accelerators, and without parsing residual RTL predicates or receiving deep architectural state prefixes, local policies still saturate before multi-step virtual-memory or coprocessor preconditions.

\textbf{LLM-Guided Hardware Verification.} Recent LLM verifiers generate testbenches, assertions, and Universal Verification Methodology (UVM) harnesses~\cite{llm4cov,assertllm,correctbench,uvllm,uvmarvel,tcad_errors}. Most closely related, FLAME~\cite{flame} invokes an LLM after random processor exploration using specification RAG, multi-turn chain-of-thought (CoT) planning, and error-guided repair, targeting RISC-V CPU verification rather than RoCC systolic or approximate block-FP accelerators. Building on post-local LLM invocation, VeriSegue closes five remaining gaps: unified CPU--accelerator adaptation via $\mathcal{A}_d$, admission-decoupled RTL CF-UCB scheduling over backward slices, sandboxed constructor synthesis beyond static menus, hybrid TraceSMT pruning and trace inversion, and bidirectional deep-state return ($\mathrm{S2}\to\mathrm{S1}$) with $K=5$ boundary batching that lets fast local rollouts explore deep microarchitectural states at zero extra cloud queries.

\section{The VeriSegue Methodology}
\label{sec:method}

VeriSegue coordinates compact on-premise RL with execution-guided reasoning from a frozen LLM in a five-step bidirectional loop (Fig.~\ref{fig:architecture}). First, on-premise Stage~1 (\S\ref{sec:method_collab}) sweeps shallow decode, pipeline, and arithmetic logic via grammar-masked local GRPO (\circnum{1}) until control-flow velocity $\gamma_{\mathrm{cf}}$ saturates at $\tau^*$, triggering an adaptive handoff of the admitted bitmap $\mathcal{C}_{\mathrm{adm}}^{(\tau^*)}$ (\circnum{2}) to Stage~2. Next, a frozen cloud or private-VPC LLM (\S\ref{sec:method_stage2}) performs admission-decoupled, stall-bounded CF-UCB scheduling, backward RTL slicing, dynamic constructor synthesis (\texttt{SynLib}), TraceSMT inversion, and $K{=}5$ boundary-diversified batching (\circnum{3}). Every Stage-1 and Stage-2 candidate traverses the shared local three-oracle harness (\S\ref{sec:method_infra}, \circnum{4}), which strictly decouples monotonic coverage admission ($\mathcal{G}_{\mathrm{cov}}{=}1 \to \mathcal{C}_{\mathrm{adm}}$) from differential bug triage ($\mathcal{D}_{\mathrm{bug}}{=}1 \to \mathcal{Q}_{\mathrm{bug}}$). Finally, verified Stage-2 witnesses return reusable deep architectural prefixes $\mathcal{S}_{\mathrm{seed}}$ (and optional KL-anchored replay windows) to Stage~1 along the outer reverse rail (\circnum{5}), unlocking deep-state local rollouts at zero additional cloud queries (Algorithm~\ref{alg:verisegue}).

\begin{algorithm}[t]
\caption{VeriSegue Closed-Loop Flow ($\mathrm{S1}\!\to\!\mathrm{S2}\!\to\!\mathrm{S1}$)}
\label{alg:verisegue}
\begin{algorithmic}[1]
\Require $\mathcal{A}_d$, $\mathcal{P}$, $\pi_{\phi}$, frozen $\Psi_{\mathrm{LLM}}$, $\mathcal{M}_d^{(0)}$, budgets $(T_{\max}, R, T_{\mathrm{return}}, T_{\mathrm{budget}})$
\Ensure Coverage $\mathcal{C}$, library $\mathcal{M}_d$, bug queue $\mathcal{Q}_{\mathrm{bug}}$
\State Init $\mathcal{C},\mathcal{Q}_{\mathrm{bug}} \gets \emptyset$, $N \gets 0$, $\mathcal{M}_d \gets \mathcal{M}_d^{(0)}$; prune $\mathcal{P}_{\mathrm{UNSAT}}$ via COI+SAT
\Repeat \Comment{\textbf{Stage 1:} Masked local GRPO}
  \State Sample $G$ masked siblings $\{x_{b,i}\}$; execute ref/RTL and gate via~(\ref{eq:gate})
  \State Score $R_b$ on frozen $(\mathcal{C}_{b-1}, N_{b-1})$ via~(\ref{eq:reward}); update $\phi$ via~(\ref{eq:grpo})
\Until{$\gamma_{\mathrm{cf}} < 10^{-5}$ ($b > W$,~(\ref{eq:dual_rate})) or $T_{\max}/t_{\max}$ reached}
\State Reset $\mathcal{T}_{\mathrm{used}} \gets \emptyset$ and module counters $(n, s, \ell) \gets 0$
\For{$r = 1 \dots R$, sims $< T_{\mathrm{budget}} - T_{\mathrm{return}}$} \Comment{\textbf{Stage 2}}
  \State Pick $m^*_r \in \mathcal{K}_r^{\mathrm{elig}}$ via CF-UCB~(\ref{eq:ucb}) ($s_{\max}=4$); build 4-hop dossier $\mathcal{D}_r$
  \State Query $\Psi_{\mathrm{LLM}}(\mathcal{D}_r, \mathcal{M}_d \setminus \mathcal{T}_{\mathrm{used}})$ for synthesis, TraceSMT, or $K=5$ batch
  \State Gate via~(\ref{eq:gate}); update $(\mathcal{C},\mathcal{Q}_{\mathrm{bug}},\mathcal{T}_{\mathrm{used}})$ and decoupled $(n,s,\ell)$; promote $\Delta\mathcal{C}>0$ into $\mathcal{M}_d$
\EndFor
\State Extract $\mathcal{S}_{\mathrm{seed}}$ and opt.\ $\mathcal{B}_{\mathrm{S2\to S1}}$; optionally adapt $\phi$ via~(\ref{eq:replay})
\State Run $T_{\mathrm{return}}$ local rollouts inside $\mathcal{S}_{\mathrm{seed}}$ ($0$ LLM calls); gate via~(\ref{eq:gate})
\State \Return $(\mathcal{C}, \mathcal{M}_d, \mathcal{Q}_{\mathrm{bug}})$
\end{algorithmic}
\end{algorithm}

\begin{figure*}[t]
  \centering
  \includegraphics[width=\textwidth]{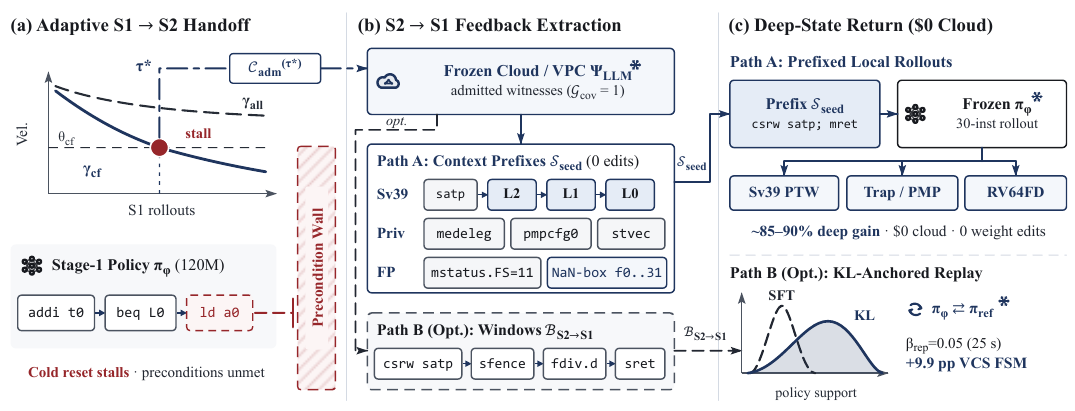}
  \caption{Adaptive $\mathrm{S1}\!\to\!\mathrm{S2}$ handoff and bidirectional $\mathrm{S2}\!\to\!\mathrm{S1}$ deep-state feedback. (a)~Cold-reset Stage~1 stalls before the precondition wall while toggle-immune $\gamma_{\mathrm{cf}}$ crosses $\theta_{\mathrm{cf}}$ at $\tau^*$, handing off $\mathcal{C}_{\mathrm{adm}}^{(\tau^*)}$. (b)~Admitted Stage-2 witnesses yield zero-retrain prefixes $\mathcal{S}_{\mathrm{seed}}$ (Path~A) and optional windows $\mathcal{B}_{\mathrm{S2\to S1}}$ (Path~B). (c)~Prepending $\mathcal{S}_{\mathrm{seed}}$ to frozen $\pi_\phi$ captures ${\sim}85\%$--$90\%$ of deep-state gain at $\$0$ cloud cost; optional KL replay ($\beta_{\mathrm{rep}}{=}0.05$) prevents SFT entropy collapse.}
  \label{fig:mech_collab}
\end{figure*}

\subsection{Instrumented Universe, Modular Adapter, and Three-Oracle Gate}
\label{sec:method_infra}

\textbf{Instrumented Coverage Universe.} Compiling an RTL design under Verilator~\cite{verilator} or VCS partitions its coverage points by metric:
\begin{equation}
  \mathcal{P} = \mathcal{P}_{\mathrm{line}} \sqcup \mathcal{P}_{\mathrm{branch}} \sqcup \mathcal{P}_{\mathrm{toggle}} \;(\sqcup\,\mathcal{P}_{\mathrm{cond}} \sqcup \mathcal{P}_{\mathrm{fsm}}),
  \label{eq:universe}
\end{equation}
where control-flow points $\mathcal{P}_{\mathrm{cf}}$ capture statement, branch, condition, and FSM transitions, while $\mathcal{P}_{\mathrm{toggle}}$ tracks register and wire bit toggles across RTL modules $m \in \mathcal{K}$. Subtracting formally proved unreachable clauses $\mathcal{P}_{\mathrm{UNSAT}}$ yields a sound SAT-pruned upper bound $\mathcal{P} \setminus \mathcal{P}_{\mathrm{UNSAT}}$ under the solver's mode assumptions. Partitioning $\mathcal{P} = \mathcal{P}_{\mathrm{real}} \sqcup \mathcal{P}_{\mathrm{instr}}$ removes testbench wrappers, clock/reset nets, and legacy FIRRTL instrumentation logic so reported coverage in~(\ref{eq:universe}) reflects genuine design behavior (Section~\ref{sec:setup}).

\textbf{Modular Per-DUT Adapter.} All target-specific interfaces reside behind a four-hook adapter $\mathcal{A}_d$ so that VeriSegue's Stage-1 learner, handoff detector, CF-UCB scheduler, and Stage-2 prompting logic transfer unchanged across scalar/OoO RISC-V processors and heterogeneous AI accelerators. Specifically, (1)~\emph{Encode/Mask} supplies the legal action grammar and token mask $m_t$; (2)~\emph{RunRef} executes the golden reference model; (3)~\emph{RunRTL+Cov} runs Verilator or VCS to extract coverage bitmaps and cycle-level traces $\tau$; and (4)~\emph{BuildFrontier} expands uncovered coverage identifiers into 4-hop backward Verilog predicates and active constructor menus.

\textbf{Decoupled Three-Oracle Gate.} Conflating coverage admission with bug detection causes two opposing errors: admitting illegal or crashing tests inflates bitmaps with spurious fault-handler toggles, whereas discarding reference--RTL discrepancies hides real defects. Let $\mathcal{O}_{\mathrm{cc}}(x), \mathcal{O}_{\mathrm{ref}}(x), \mathcal{O}_{\mathrm{rtl}}(x), \mathcal{E}_{\mathrm{diff}}(x) \in \{0,1\}$ denote compilation validity, reference completion, clean RTL exit within $C_{\max}$ cycles, and architectural payload divergence for candidate $x$. VeriSegue decouples coverage admission ($\mathcal{G}_{\mathrm{cov}}$) and constructor promotion ($\mathcal{G}_{\mathrm{prom}}$) from differential bug triage ($\mathcal{D}_{\mathrm{bug}}$):
\begin{equation}
\begin{split}
  \mathcal{G}_{\mathrm{cov}}(x) &= \mathcal{O}_{\mathrm{cc}}(x)\,\mathcal{O}_{\mathrm{ref}}(x)\,\mathcal{O}_{\mathrm{rtl}}(x)\bigl(1 - \mathcal{E}_{\mathrm{diff}}(x)\bigr), \\
  \mathcal{G}_{\mathrm{prom}}(x) &= \mathcal{G}_{\mathrm{cov}}(x)\,\mathbb{I}\!\left[\Delta\mathcal{C}(x) > 0\right], \\
  \mathcal{D}_{\mathrm{bug}}(x) &= \mathcal{O}_{\mathrm{cc}}(x)\Bigl(\mathbb{I}\!\left[\mathcal{O}_{\mathrm{ref}}(x) \neq \mathcal{O}_{\mathrm{rtl}}(x)\right] \\
  &\quad + \mathcal{O}_{\mathrm{ref}}(x)\,\mathcal{O}_{\mathrm{rtl}}(x)\,\mathcal{E}_{\mathrm{diff}}(x)\Bigr).
\end{split}
\label{eq:gate}
\end{equation}
Uncompiled candidates and mutual timeouts are discarded; crash asymmetries ($\mathcal{O}_{\mathrm{ref}}\neq\mathcal{O}_{\mathrm{rtl}}$) and functional divergences ($\mathcal{E}_{\mathrm{diff}}=1$) route exclusively to $\mathcal{Q}_{\mathrm{bug}}$ with $\mathcal{G}_{\mathrm{cov}}=\mathcal{G}_{\mathrm{prom}}=0$; and only consistent valid runs update the monotonic archive ($\mathcal{C}_{b-1}\subseteq\mathcal{C}_b\subseteq\mathcal{P}\setminus\mathcal{P}_{\mathrm{UNSAT}}$) and qualify for \texttt{SynLib} promotion. Importantly, triage distinguishes an implementation-matched simulator (such as AxCore's shipped FP32 CUDA model, which replicates hardware wraparounds) from an independent intent-level oracle (Spike or an exact rational block-FP oracle).

\subsection{Coverage-Adaptive Local--Cloud Collaboration and Deep-State Return}
\label{sec:method_collab}

\textbf{Grammar-Masked, Critic-Free Local Exploration (Stage~1).} Stage~1 harvests shallow decode, pipeline-hazard, arithmetic logic unit (ALU), and cache coverage on a single workstation GPU at zero cloud cost. Prior local RL generators such as GenHuzz~\cite{genhuzz} encounter two bottlenecks on bare-metal RISC-V binaries. First, narrow vocabularies ($2{,}421$ symbols omitting branch and synchronization instructions) and unmasked sampling reduce $L=30$-instruction bare-metal compilation to $0\%$ under both Verilator and VCS. Second, once shallow logic is swept, per-test novelty falls by $>99.8\%$, destabilizing actor--critic PPO value heads (producing negative late EV of $-0.05$ to $-0.42$ while doubling BF16 GPU memory).

\begin{figure*}[t]
  \centering
  \includegraphics[width=\textwidth]{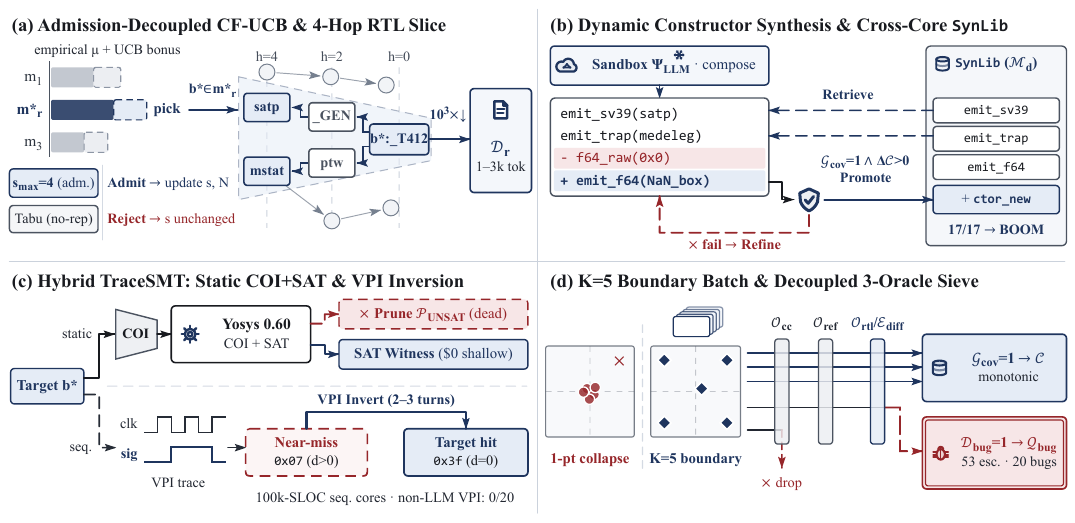}
  \caption{Execution-guided Stage-2 frontier closure under frozen $\Psi_{\mathrm{LLM}}$ (snowflake). (a)~Admission-decoupled CF-UCB ($s_{\max}{=}4$, tabu) selects target $b^*$ and compresses $>100\mathrm{k}$-SLOC RTL via 4-hop backward slicing into a $1$--$3\mathrm{k}$-token dossier $\mathcal{D}_r$. (b)~Sandboxed \texttt{SynLib} promotes \texttt{ctor\_new} iff $\mathcal{G}_{\mathrm{cov}}{=}1 \wedge \Delta\mathcal{C}{>}0$ with multi-turn self-repair. (c)~Hybrid TraceSMT pairs static COI+SAT pruning with $2$--$3$-turn VPI near-miss inversion. (d)~$K{=}5$ boundary batching feeds the three-oracle gate separating $\mathcal{C}$ from $\mathcal{Q}_{\mathrm{bug}}$ ($53$ escapes, $20$ root causes).}
  \label{fig:mech_stage2}
\end{figure*}

We resolve both bottlenecks using a compact $120\mathrm{M}$-parameter ISA-GLM~\cite{glm} policy $\pi_\phi$ over an $11{,}642$-symbol RV64GC vocabulary generating $L=30$-instruction sequences (matching bare-metal basic-block horizons). At each decoding step $t$ given prefix state $s_t$, $\mathcal{A}_d.\mathrm{Encode/Mask}$ zeroes syntactically illegal logits before softmax to produce a masked action distribution $\tilde{\pi}_\phi(a_t \mid s_t, m_t)$, raising $L=30$ compilation from $0\%$ to ${\sim}100\%$. To remove worker-order bias across parallel simulators, every test $x$ in batch $b$ is scored against the frozen batch-entry snapshot $(\mathcal{C}_{b-1}, N_{b-1})$, where $N_{b-1}$ stores cumulative per-point hit counts. The reward combines weighted coverage density $c_{\mathrm{w}}(x)$ ($70\%$ control-flow, $30\%$ all points), log newly covered control-flow and toggle counts ($\Delta_b^{\mathrm{cf}}, \Delta_b^{\mathrm{tog}}$), and inverse-frequency control-flow rarity $r_b^{\mathrm{rare}}(x)$:
\begin{equation}
  R_b(x) = \begin{cases}
    \alpha_1 c_{\mathrm{w}}(x) + \alpha_2\ln(1{+}\Delta_b^{\mathrm{cf}}) \\
    +\,\alpha_3\ln(1{+}\Delta_b^{\mathrm{tog}}) + \alpha_4 r_b^{\mathrm{rare}}, & \mathcal{G}_{\mathrm{cov}}{=}1, \\[3pt]
    -\alpha_0, & \mathcal{G}_{\mathrm{cov}}{=}0,
  \end{cases}
  \label{eq:reward}
\end{equation}
with $(\alpha_1,\alpha_2,\alpha_3,\alpha_4,\alpha_0)=(6.0,8.0,1.5,4.0,0.5)$. Here $\alpha_2 > \alpha_3$ offsets the $6\times$--$12\times$ numerical dominance of toggle points over control-flow points, while Section~\ref{sec:main_eval} confirms low sensitivity to reward shaping. Removing the unstable value critic, Stage~1 updates $\phi$ via critic-free GRPO~\cite{grpo,deepseekmath} across $G=8$ sibling programs per group. Group-normalized advantages $\hat{A}_{b,i} = (R_b(x_{b,i}) - \mu_b)/(\sigma_b + 10^{-6})$ (using group mean $\mu_b$ and standard deviation $\sigma_b$) drive a Kullback--Leibler (KL) regularized asymmetric clipped objective~\cite{dapo} with importance ratio $\rho_{i,t} = \tilde{\pi}_{\phi}/\tilde{\pi}_{\phi_{\mathrm{old}}}$, clip $c_{i,t} = \mathrm{clip}(\rho_{i,t}, 1-\epsilon_{\mathrm{low}}, 1+\epsilon_{\mathrm{high}})$ ($\epsilon_{\mathrm{low}}=0.20, \epsilon_{\mathrm{high}}=0.28$, encouraging exploration under sparse positive advantages), reference policy $\tilde{\pi}_{\mathrm{ref}}$, and $\beta=10^{-2}$:
\begin{equation}
\begin{split}
  \mathcal{J}_{\mathrm{GRPO}}(\phi) = \mathbb{E}_{i,t}\Bigl[&\min\bigl(\rho_{i,t}\hat{A}_{b,i},\, c_{i,t}\hat{A}_{b,i}\bigr) \\
  &- \beta\,\widehat{D}_{\mathrm{KL}}\bigl(\tilde{\pi}_{\phi}\,\|\,\tilde{\pi}_{\mathrm{ref}}\bigr)\Bigr].
\end{split}
\label{eq:grpo}
\end{equation}
Together,~(\ref{eq:reward}) and~(\ref{eq:grpo}) reduce peak BF16 VRAM by $34.8\%$--$59.7\%$ across $120\mathrm{M}$--$1.3\mathrm{B}$ backbones and run $1.96\times$ faster than masked GenHuzz PPO on VCS while improving all five coverage metrics.

\textbf{Toggle-Immune Control-Flow Velocity Handoff.} Selecting when to hand off from local RL to cloud reasoning involves a direct trade-off (Fig.~\ref{fig:mech_collab}(a)). Fixed simulation splits waste local cycles on small accelerators or trigger prematurely on deep processor pipelines, whereas aggregate coverage velocity ($\gamma_{\mathrm{all}}$) stays artificially positive from random datapath bit flips long after control-flow discovery stalls. Stage~1 therefore tracks control-flow discovery velocity over a trailing $W=20$-batch smoothing window (chosen to filter single-batch noise without delaying handoff):
\begin{equation}
  \gamma_{\mathrm{cf}}(b; W) = \frac{|\mathcal{C}_b \cap \mathcal{P}_{\mathrm{cf}}| - |\mathcal{C}_{b-W} \cap \mathcal{P}_{\mathrm{cf}}|}{\Delta T_W \cdot |\mathcal{P}_{\mathrm{cf}}|},
  \label{eq:dual_rate}
\end{equation}
where $\Delta T_W$ is the number of simulations executed in window $W$. Stage~1 halts at batch $\tau^*$, defined as the earliest point where $\gamma_{\mathrm{cf}} < 10^{-5}$ for $b > W$ in~(\ref{eq:dual_rate}), or where simulation cap $T_{\max}$ or wall-clock cap $t_{\max}$ is reached. Because $\gamma_{\mathrm{cf}}$ isolates control-flow saturation, it transfers consistently between Verilator branches/lines and VCS conditions/FSMs and remains invariant across a $5\times$ threshold sweep. At handoff, only the admitted coverage bitmap $\mathcal{C}_{\mathrm{adm}}^{(\tau^*)}$ passes to the local Stage-2 frontier builder.

\textbf{Bidirectional Deep-State Context Transfer ($\mathrm{S1} \to \mathrm{S2} \to \mathrm{S1}$).} One-way LLM hole-closure pipelines discard the architectural setup constructed during Stage~2, forcing subsequent local exploration to restart from blank reset and stall again before the precondition wall (Fig.~\ref{fig:mech_collab}(a)--(c)). VeriSegue closes this reverse loop while keeping Stage-2 LLM weights frozen. From verified Stage-2 witnesses, we extract reusable \emph{deep architectural initialization contexts} $\mathcal{S}_{\mathrm{seed}}$ (Fig.~\ref{fig:mech_collab}(b)): (i)~coherent 3-level Sv39 page tables, active \texttt{satp} mappings, and supervisor trap delegation (\texttt{medeleg}/\texttt{mideleg}); (ii)~width-matched RV64FD floating-point register states preventing silent NaN-boxing faults; and (iii)~active Physical Memory Protection (PMP) and exception-vector setups. We also collect optional high-gain instruction windows $\mathcal{B}_{\mathrm{S2\to S1}}$.

During the return pass ($T_{\mathrm{return}}$ local rollouts, Fig.~\ref{fig:mech_collab}(c)), Path~A prepends $\mathcal{S}_{\mathrm{seed}}$ directly so the frozen compact policy $\pi_\phi$ runs syntax-masked rollouts inside deep microarchitectural states at local simulation speed and zero extra cloud queries. On extended return budgets, optional Path~B adapts $\phi$ via $25\,\mathrm{s}$ of 2-epoch KL-anchored supervised replay over instruction windows $w \in \mathcal{B} \triangleq \mathcal{B}_{\mathrm{S2\to S1}}$ of length $|w|$:
\begin{equation}
\begin{split}
  \mathcal{L}_{\mathrm{replay}}(\phi) = \mathbb{E}_{w \sim \mathcal{B}}\Bigl[&-\tfrac{1}{|w|}\sum_t \log\tilde{\pi}_\phi(a_{w,t}\!\mid\! s_{w,t}, m_{w,t}) \\
  &+ \beta_{\mathrm{rep}}\,\widehat{D}_{\mathrm{KL}}\bigl(\tilde{\pi}_{\mathrm{ref}}\,\|\,\tilde{\pi}_\phi\bigr)\Bigr].
\end{split}
\label{eq:replay}
\end{equation}
In~(\ref{eq:replay}), $\beta_{\mathrm{rep}}=0.05$ is set $5\times$ stronger than Stage-1 $\beta$ to anchor policy entropy against the collapse observed under un-anchored fine-tuning on compact witness buffers.

\subsection{Execution-Guided Stage-2 Frontier Closure}
\label{sec:method_stage2}

Following handoff of $\mathcal{C}_{\mathrm{adm}}^{(\tau^*)}$, Stage~2 drives a frozen LLM $\Psi_{\mathrm{LLM}}$ (zero LLM fine-tuning) through four execution-grounded mechanisms (Fig.~\ref{fig:mech_stage2}) orchestrated by Algorithm~\ref{alg:verisegue}.

\begin{table*}[t]
\centering
\caption{VCS-MX five-metric evaluation on RocketTile (\texttt{-cm line+cond+fsm+tgl+branch}, \texttt{dut\_inst}). Panels~(a)--(c) report native Stage-1 training, live online scheduling, and \texttt{urg} VDB merges ($n=3$, $C_{\max}=6{,}000$; DS=\texttt{DeepSeek-V4.1-Flash}, QW=\texttt{Qwen3.8-Max}). Panel~(d) cross-validates all four $60\,\mathrm{min}$ equal-horizon arms ($n=5$ seeds, $1{,}688$ unique RV64GC programs replayed on \texttt{simv\_rocket\_fsm}, $C_{\max}=16{,}000$): VeriSegue (Full) wins $5/5$ seeds over FLAME across all six VCS metrics (\texttt{mod\_def}: $63.29\%$ Score, $36.58\%$ FSM).}
\label{tab:vcs_native}
\footnotesize
\setlength{\tabcolsep}{0pt}
\renewcommand{\arraystretch}{1.03}
\begin{tabular*}{\textwidth}{@{\extracolsep{\fill}}lrrrrrrl@{}}
\toprule
\textbf{VCS Protocol / Arm} & \textbf{Score} & \textbf{Line} & \textbf{Cond} & \textbf{FSM (Hit)} & \textbf{Br.} & \textbf{Tgl.} & \textbf{Notes / Wall / Stats} \\
\midrule
\multicolumn{8}{@{}l@{}}{\textit{(a) Native VCS Stage-1 Online Training ($960$ tests/run, $n=3$, seeds $\{613,719,823\}$; all metrics in \%)}} \\
GenHuzz PPO (Unmasked, C+L+F) & $0.00$ & $0.00$ & $0.00$ & $0.00$ ($0.0$) & $0.00$ & $0.00$ & $0\%$ adm.\ ($29.0\,\mathrm{m}$); mask req. \\
GenHuzz PPO (Masked, C+L+F) & $46.51$ & $56.18$ & $72.85$ & $19.14$ ($20.7$) & $59.25$ & $25.15$ & Late EV $-0.09$..$-0.19$; $38.5\,\mathrm{m}$ \\
\textbf{VeriSegue Stage-1 GRPO (Masked)} & $\mathbf{47.25}$ & $\mathbf{56.62}$ & $\mathbf{73.44}$ & $\mathbf{19.44}$ ($21.0$) & $\mathbf{59.62}$ & $\mathbf{27.15}$ & All 5 higher; $19.6\,\mathrm{m}$ ($1.96\times$) \\
\midrule
\multicolumn{8}{@{}l@{}}{\textit{(b) Live Online VCS Closed-Loop Scheduling ($16$ rounds over $28$ constructors, $n=3$)}} \\
Live VCS Loop: Uniform Random & $45.51$ & $54.43$ & $72.41$ & $17.90$ ($19.3$) & $58.68$ & $24.13$ & High SD ($\pm 626$ L / $\pm 363$ C) \\
\textbf{Live VCS Loop: CF-UCB ($s_{\max}=4$)} & $\mathbf{49.96}$ & $\mathbf{58.73}$ & $\mathbf{77.23}$ & $\mathbf{26.85}$ ($29.0$) & $\mathbf{62.15}$ & $\mathbf{24.84}$ & $+8.95\,$pp FSM; $20$--$66\times$ SD$\downarrow$ \\
\midrule
\multicolumn{8}{@{}l@{}}{\textit{(c) Exact \texttt{urg} VDB End-to-End Merges (Stage~1 vs.\ Closed-Loop $\mathrm{S1} \to \mathrm{S2} \to \mathrm{S1}$, $n=3$)}} \\
$\mathrm{S1}$ Only (VeriSegue GRPO, $960$ tests) & $47.25$ & $56.62$ & $73.44$ & $19.44$ ($21.0$) & $59.62$ & $27.15$ & Cold ceiling (Union $19.44\%$) \\
$\mathrm{S1}\!\to\!\mathrm{S2}\!\to\!\mathrm{S1}$ VeriSegue Frozen (DS/QW) & $54.5$/$56.0$ & $60.7$/$61.5$ & $81.0$/$83.7$ & $31.5$/$32.7$ & $64.8$/$66.7$ & $34.3$/$35.2$ & $0$-retrain (Union $34.3\%$) \\
\textbf{$\mathrm{S1}\!\to\!\mathrm{S2}\!\to\!\mathrm{S1}$ VeriSegue + KL-Rep (QW)} & $\mathbf{56.17}$ & $\mathbf{61.52}$ & $\mathbf{83.88}$ & $\mathbf{33.03}$ ($35.7$) & $\mathbf{66.93}$ & $\mathbf{35.49}$ & $+13.59\,$pp FSM / $+10.44\,$pp C \\
\midrule
\multicolumn{8}{@{}l@{}}{\textit{(d) Equal-Horizon ($60\,\mathrm{min}$) Four-Arm FLAME Replay on VCS ($n=5$, \texttt{dut\_inst})}} \\
GenHuzz (Local RL Only, Sat.) & $45.62$ & $52.25$ & $70.89$ & $22.22$ ($24.0$) & $57.18$ & $25.57$ & Local ceiling ($60.0\,\mathrm{m}$, $\$0$) \\
FLAME & $48.61$ & $55.80$ & $75.03$ & $24.26$ ($26.2$) & $60.32$ & $27.63$ & $200$ calls ($60.0\,\mathrm{m}$, $\$0.200$) \\
VeriSegue (No Deep-State Return) & $55.18$ & $60.83$ & $84.70$ & $30.74$ ($33.2$) & $66.31$ & $33.34$ & $69.6$ calls ($60.0\,\mathrm{m}$, $\$0.333$) \\
\textbf{VeriSegue (Full Bidirectional)} & $\mathbf{57.78}$ & $\mathbf{61.93}$ & $\mathbf{88.24}$ & $\mathbf{32.04}$ ($\mathbf{34.6}$) & $\mathbf{68.80}$ & $\mathbf{37.87}$ & $5/5$ ($p{\le}0.024$) vs.\ FLAME \\
\bottomrule
\end{tabular*}
\end{table*}

\textbf{1.\ Stall-Bounded CF-UCB and 4-Hop Residual RTL Dossiers (Fig.~\ref{fig:mech_stage2}(a)).} Feeding raw generated RTL ($>100\mathrm{k}$ lines) or module-level percentages to an LLM fails because Chisel/SpinalHDL netlists use compiler-generated signals (\texttt{\_T\_412}, \texttt{\_GEN\_17}) that obscure architectural intent, while naive prompting stalls on hard modules. Each round $r$, CF-UCB selects an eligible RTL module $m \in \mathcal{K}_r^{\mathrm{elig}}$ balancing size-damped residual control-flow density $q_{r,m}$, exploration over selection counts $n_{r,m}$, and stall damping $s_{r,m}$:
\begin{equation}
\begin{split}
  q_{r,m} &= \frac{u_{r-1,m}}{|\mathcal{P}_m|} + w_{\mathrm{cf}}\,\frac{u_{r-1,m}^{\mathrm{cf}}}{\max(\kappa_{\mathrm{cf}}, |\mathcal{P}_m^{\mathrm{cf}}|)}, \\
  \mathrm{UCB}_r(m) &= q_{r,m} + c\sqrt{\frac{\ln(r + 1)}{n_{r,m} + 1}} - \lambda\,s_{r,m},
\end{split}
\label{eq:ucb}
\end{equation}
where $u_{r-1,m}$ and $u_{r-1,m}^{\mathrm{cf}}$ denote total and control-flow uncovered points in module $m$, and $|\mathcal{P}_m|$ and $|\mathcal{P}_m^{\mathrm{cf}}|$ are the module's total and control-flow point counts. We fix $(w_{\mathrm{cf}}, \kappa_{\mathrm{cf}}, c, \lambda)=(0.92, 250, 1.43, 0.80)$ once from preliminary calibration and hold all four values frozen across all five designs (including unseen CVA6); here $c \approx \sqrt{2}$ follows standard UCB scaling, while $\kappa_{\mathrm{cf}}=250$ prevents tiny wrapper modules from dominating scheduling. To prevent module starvation under linear stall damping alone, we enforce a consecutive-stall cooldown cap $s_{\max}=4$ (resetting only when all active modules saturate while excluding $m^*_{r-1}$) alongside without-replacement constructor tabu $\mathcal{T}_{\mathrm{used}}$. Crucially, stall updates are decoupled from syntax or admission rejects (leaving $s_{r,m}$ unchanged on rejected drafts, Fig.~\ref{fig:mech_stage2}(a)) so malformed trials never prematurely cool down valid deep modules, while scoring all non-wrapper RTL modules dynamically alongside uncovered functional bins and missing VCS FSM arcs eliminates static subsystem blind spots. For $m^*_r$, $\mathcal{A}_d.\mathrm{BuildFrontier}$ extracts concrete Verilog \texttt{if}-branch predicates and walks up to $4$ hops backward along the combinational assignment graph to driving control registers, opcodes, or operands, compressing $>100\mathrm{k}$-SLOC RTL by $>1{,}000\times$ into a $1.2\mathrm{k}$--$2.9\mathrm{k}$-token dossier $\mathcal{D}_r$.

\textbf{2.\ Dynamic Constructor Synthesis (\texttt{SynLib}) and Closed-Loop Self-Repair (Fig.~\ref{fig:mech_stage2}(b)).} Fixed human-authored constructor menus impose a behavioral ceiling across models. When existing menu entries saturate, $\Psi_{\mathrm{LLM}}$ synthesizes new parameterized constructors by composing verified architectural micro-primitives inside a restricted sandbox enforcing static syntax, register-safety, and termination bounds. Failed or low-gain drafts undergo bounded multi-turn execution-guided self-repair using exact assembler, reference-model, and differential diagnostics, while compositional multi-cycle $\mathrm{S1}\leftrightarrow\mathrm{S2}$ return stacks verified deep contexts sequentially. A synthesized constructor enters persistent library \texttt{SynLib} ($\mathcal{M}_d$) if and only if it passes~(\ref{eq:gate}) ($\mathcal{G}_{\mathrm{cov}}=1$) and adds positive newly admitted coverage ($\Delta\mathcal{C} > 0$). Because promoted constructors encode reusable ISA-level preconditions rather than core-specific workarounds, promoted constructors transfer zero-shot across distinct microarchitectures (Section~\ref{sec:deepstate_eval}).

\textbf{3.\ TraceSMT: Static COI+SAT Pruning and Near-Miss Inversion (Fig.~\ref{fig:mech_stage2}(c)).} Sequential BMC from reset blows up on $100\mathrm{k}$-line multi-cycle processors and $64$-row systolic arrays, while static Verilog prompting fails when arithmetic datapaths hide internal offsets. TraceSMT decouples formal pruning from sequential search into two lightweight tiers. First, \emph{static COI+SAT pruning and shallow solving} runs Yosys~$0.60$~\cite{yosys} locally on residual clauses, certifying unreachable clauses UNSAT with zero false eliminations and resolving shallow 1-DFF bit-level targets at $\$0$ cloud cost. Second, for multi-cycle sequential targets beyond SAT, \emph{dynamic near-miss trace inversion} logs the closest-cycle Verilog Procedural Interface (VPI) signal watermark $x_k$ and scalar distance $d(x_k, g) \ge 0$ for SAT-feasible targets $g$, returning concrete numeric feedback so $\Psi_{\mathrm{LLM}}$ can invert hidden guards or arithmetic offsets in $2$--$3$ turns (e.g., turning \texttt{max\_pageValid\_mask}=\texttt{0x07} into \texttt{0x3f} on RocketChip or inverting PreAdd's $+172$ alignment bias on AxCore).

\begin{figure*}[t]
  \centering
  \includegraphics[width=\textwidth]{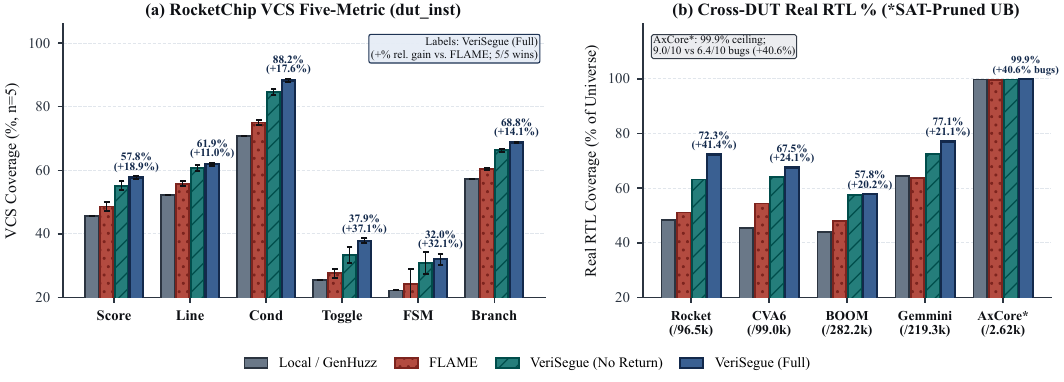}
  \caption{Dual-simulator four-arm panorama ($n=5$ seeds, mean $\pm$ $95\%$ CI, y-axis from $20\%$ with relative gain vs.\ FLAME annotated). (a)~VCS-MX five-metric replay on RocketTile (\texttt{dut\_inst}, $60\,\mathrm{min}$ horizon): VeriSegue (Full) wins $5/5$ seeds over FLAME across all six metrics. (b)~Equal-horizon real RTL coverage percentage across all five designs under Verilator exact denominators ($60\,\mathrm{min}$ CPUs, $200\,\mathrm{min}$ Gemmini; AxCore* over its $2{,}621$-point SAT-pruned upper bound, where VeriSegue hits $99.89\%$ and exposes $9.0/10$ vs.\ $6.4/10$ defects).}
  \label{fig:vcs_panorama}
\end{figure*}

\textbf{4.\ Boundary-Diversified Batching and Decoupled Triage (Fig.~\ref{fig:mech_stage2}(d)).} Single-point LLM prompting suffers from greedy parameter collapse, repeatedly emitting nominal defaults (such as default rounding mode or unit tile dimension $\texttt{dim\_J}=1$) that compile cleanly while avoiding zero-dimension, saturation, and signed-boundary faults. To counter mode collapse without defect hints, Stage~2 emits $K=5$ candidates per round spanning five boundary regions (minimum/zero, interior, mode contrast, upper saturation, and edge encoding), verified by a schema-only diversity check ($\le 1$ syntax repair) and evaluated through~(\ref{eq:gate}).

\textbf{Closed-Loop Budget Accounting.} As summarized in Algorithm~\ref{alg:verisegue}, our single-return-pass workflow reserves $T_{\mathrm{return}}$ local simulations for deep-state exploration and terminates Stage~2 at the earliest of round cap $R$, simulation budget $T_{\mathrm{budget}} - T_{\mathrm{return}}$ (charging every candidate in a $K=5$ batch, every syntax repair, and every TraceSMT refinement turn), or frontier exhaustion. Total cloud queries are bounded by $Q_{\mathrm{LLM}} \le R(1 + r_{\mathrm{rep}}) + N_{\mathrm{ref}}T_{\mathrm{ref}}$, where $r_{\mathrm{rep}} \le 1$ caps schema repairs ($r_{\mathrm{rep}} \le 3$ for free-form constructor synthesis) and $N_{\mathrm{ref}}$ targeted clauses use at most $T_{\mathrm{ref}}$ refinement turns each.

\section{Experimental Evaluation}
\label{sec:main_eval}

We evaluate VeriSegue across five open-source RTL designs and two simulators along five axes: setup and dual-simulator protocol (Section~\ref{sec:setup}), VCS five-metric validation (Section~\ref{sec:vcs_eval}), five-seed full-horizon comparisons against FLAME across all five targets (Section~\ref{sec:e2e_eval}), Stage-1 RL isolation (Section~\ref{sec:s1_eval}), and handoff, slicing, counterfactual, and component ablations (Section~\ref{sec:ablation_cost}).

\subsection{Experimental Setup and Dual-Simulator Protocol}
\label{sec:setup}

\textbf{Five RTL Benchmarks and Exact Coverage Universes.} Our benchmark suite spans three stimulus domains across five open-source hardware designs. \emph{RocketChip}~\cite{rocketchip} (scalar RV64GC) has $47$ modules, $139{,}443$ Verilator points ($96{,}515$ non-DIFUZZRTL real RTL points used as our $100\%$ denominator: $2{,}879$ Line, $9{,}974$ Branch, and $85{,}249$ real Toggle), $28$ constructors ($461$ SLOC), and a $299$-SLOC harness; under VCS (\texttt{simv\_rocket\_fsm}, \texttt{dut\_inst}/\texttt{mod\_def}), RocketTile exposes $14{,}627$/$12{,}841$ Line, $6{,}444$/$5{,}569$ Cond, $108$/$111$ FSM, $17{,}418$/$15{,}040$ Branch, and $660{,}480$/$567{,}988$ Toggle points. \emph{Unseen SystemVerilog CVA6 / Ariane}~\cite{cva6} (\texttt{CVA6Config.covO0}) has $133$ modules and $99{,}033$ real RTL points ($99{,}234$ total: $1{,}740$ Line, $1{,}990$ Branch, $95{,}504$ Toggle), evaluated with zero core verification edits and $0\,\text{GPU-h}$ RL retraining behind a lightweight bare-metal harness (${\sim}120$ SLOC assembly init plus $379$ SLOC simulator/Spike glue and $326$ SLOC RPC transport). \emph{SonicBOOM}~\cite{sonicboom} (\texttt{SmallBoomTile}, superscalar OoO RV64GC) has $64$ modules and $282{,}177$ points ($15{,}415$ Line, $25{,}426$ Branch, $241{,}336$ Toggle). \emph{Gemmini}~\cite{gemmini} ($16{\times}16$ C+RoCC systolic array in Chipyard~\cite{chipyard}) has $46$ modules and $219{,}285$ Gemmini RTL points ($1{,}055$ Line, $15{,}254$ Branch, $202{,}976$ Toggle). \emph{AxCore}~\cite{axcore} (approximate block-FP matrix engine) has $9$ modules, $2{,}708$ nominal points ($27$ Line, $34$ Branch, $61$ Expr, $2{,}586$ Toggle), and an $82$-SLOC exact rational oracle; Yosys~$0.60$ COI+SAT certifies $87$ points UNSAT ($46$ globally dead RTL bits, $38$ FP4-mode clauses, $3$ static testbench pins), establishing a $2{,}621$-point sound SAT-pruned upper bound (non-refuted universe, rather than a claim that all remaining points are dynamically reachable from reset). Under VCS (\texttt{vcs\_axcore/simv}), AxCore's $13$ core datapath modules expose $37$ Line, $37$ Cond, $58$ Branch, $4{,}518$ Toggle, and $0$ FSM points (pure datapath).

\textbf{Unified Functional Covergroup and Instrumentation Audit.} Alongside real RTL coverage, all three RV64GC processors share an identical $96$-bin architectural functional covergroup spanning six deep-state subsystems (\texttt{MMU\_PTW\_TLB} $18$, \texttt{CSR\_TRAP\_PMP} $24$, \texttt{FPU} $12$, \texttt{MULDIV\_AMO} $12$, \texttt{FRONTEND\_CACHE} $12$, \texttt{CROSS\_DEEP} $18$), while Gemmini uses a $64$-bin RoCC/CISC functional model and AxCore tracks $50$ arithmetic regime bins plus $10$ canonical in-envelope RTL defects (\texttt{AX-D1}--\texttt{D10}). Stripping wrapper, clock/reset, and legacy FIRRTL nets leaves $92.0\%$--$99.9\%$ pure design RTL across all five designs ($92{,}553$-point architectural mask on RocketChip for deep-state attribution). Pre-place-and-route (pre-PnR) $45\,\mathrm{nm}$ AxCore synthesis uses \nangate{}~\cite{nangate45} ($1.10\,\mathrm{V}$, $25\,^\circ\mathrm{C}$) via Yosys~$0.60$ and OpenSTA~$2.7.0$~\cite{opensta}.

\begin{table*}[t]
\centering
\caption{Four-arm, five-seed equal-horizon comparison against FLAME (\texttt{superpung/flame}, adapted for accelerators; FLAME: \texttt{DeepSeek-V4.1-Flash}; VeriSegue: frozen \texttt{DeepSeek-V4.1-Flash}, \texttt{Qwen3.8-Max}, and \texttt{GLM-5.2}/\texttt{5.3}) across all five RTL designs ($60.0\,\mathrm{min}$ across all three RISC-V processors; $200.0\,\mathrm{min}$ on Gemmini; $T=36$ tiles on AxCore). VeriSegue (Full) achieves unanimous $5/5$ wins ($MW_p=0.0079$, $\delta=+1.00$) over FLAME on real RTL ($\%$), Line, Branch, Toggle, and functional coverage across all five targets while sweeping $7.0/7$ Gemmini defects (plus live 0-day \texttt{GEM-D8}) and $9.0/10$ AxCore defects.}
\label{tab:e2e_eval}
\scriptsize
\setlength{\tabcolsep}{2.0pt}
\renewcommand{\arraystretch}{1.00}
\begin{tabular*}{\textwidth}{@{\extracolsep{\fill}}l rrrrrrrl@{}}
\toprule
\textbf{DUT / System Arm ($n=5$)} & \textbf{Real RTL (pts / \%)} & \textbf{Rel.\ $\Delta$} & \textbf{Line} & \textbf{Br.} & \textbf{Tgl.} & \textbf{Func (\%)} & \textbf{Wall / Cost} & \textbf{Key Metric / Defects} \\
\midrule
\multicolumn{9}{@{}l@{}}{\textit{(a) RocketChip RV64GC ($60.0\,\mathrm{min}$ equal horizon, $96{,}515$ real RTL; L/B/T in \%)}} \\
GenHuzz (Local RL, Sat.) & $46{,}535$ ($48.22\%$) & --- & $62.4$ & $39.7$ & $48.3$ & $42.7 \pm 0.0$ & $60.0\mathrm{m}$ / $\$0$ & MMU $3.5\mathrm{k}$; $0/5\ge 60\mathrm{k}$ \\
FLAME & $49{,}330$ ($51.11\%$) & $0.0\%$ & $67.0$ & $43.6$ & $50.9$ & $96.7 \pm 0.9$ & $60.0\mathrm{m}$ / $\$0.200$ & MMU $4.3\mathrm{k}$; $40.1\%$ adm. \\
VeriSegue (No Return) & $60{,}952$ ($63.15\%$) & $+23.6\%$ & $70.7$ & $51.1$ & $63.1$ & $95.8 \pm 5.3$ & $60.0\mathrm{m}$ / $\$0.333$ & MMU $8.6\mathrm{k}$; $3/5\ge 60\mathrm{k}$ \\
\textbf{VeriSegue (Full)} & $\mathbf{69{,}756}$ ($\mathbf{72.27\%}$) & $\mathbf{+41.4\%}$ & $\mathbf{71.0}$ & $\mathbf{53.2}$ & $\mathbf{72.8}$ & $\mathbf{99.6 \pm 0.9}$ & $60.0\mathrm{m}$ / $\$0.286^\dagger$ & \textbf{MMU $12.0\mathrm{k}$ ($2.79\times$); $5/5{\ge}65\mathrm{k}$} \\
\midrule
\multicolumn{9}{@{}l@{}}{\textit{(b) Unseen SystemVerilog CVA6 Zero-Shot ($60.0\,\mathrm{min}$ equal horizon, $99{,}033$ real RTL, $0\,\text{GPU-h}$ RL retraining)}} \\
GenHuzz (Zero-Shot, Sat.) & $44{,}941$ ($45.38\%$) & --- & $55.6$ & $60.7$ & $45.0$ & $38.5 \pm 1.8$ & $60.0\mathrm{m}$ / $\$0$ & MMU $4.4\%$; $5/5$ D1 \\
FLAME & $53{,}875$ ($54.40\%$) & $0.0\%$ & $60.3$ & $67.2$ & $54.1$ & $75.8 \pm 11.1$ & $60.0\mathrm{m}$ / $\$0.146$ & MMU $27.8\%$; $2/5$ D1 \\
VeriSegue (No Return) & $63{,}565$ ($64.19\%$) & $+18.0\%$ & $66.2$ & $73.0$ & $64.0$ & $\mathbf{100.0 \pm 0.0}$ & $60.0\mathrm{m}$ / $\$0.127$ & MMU $100\%$; $0/5\ge 65\mathrm{k}$ \\
\textbf{VeriSegue (Full)} & $\mathbf{66{,}855}$ ($\mathbf{67.51\%}$) & $\mathbf{+24.1\%}$ & $\mathbf{66.3}$ & $\mathbf{73.3}$ & $\mathbf{67.5}$ & $\mathbf{100.0 \pm 0.0}$ & $60.0\mathrm{m}$ / $\mathbf{\$0.124}$ & \textbf{MMU $100\%$; $5/5{\ge}65\mathrm{k}$ ($1.18\times\$$)} \\
\midrule
\multicolumn{9}{@{}l@{}}{\textit{(c) Superscalar OoO SonicBOOM ($60.0\,\mathrm{min}$ equal horizon, $282{,}177$ real RTL)}} \\
GenHuzz (Local RL, Sat.) & $123{,}955$ ($43.93\%$) & --- & $51.1$ & $52.5$ & $42.6$ & $41.7 \pm 0.0$ & $60.0\mathrm{m}$ / $\$0$ & MMU $8.0\%$; FPU $43.8\%$ \\
FLAME & $135{,}737$ ($48.10\%$) & $0.0\%$ & $55.4$ & $56.2$ & $46.8$ & $86.0 \pm 5.8$ & $60.0\mathrm{m}$ / $\$0.130$ & MMU $10.8\%$; FPU $48.1\%$ \\
VeriSegue (No Return) & $162{,}619$ ($57.63\%$) & $+19.8\%$ & $60.6$ & $62.1$ & $57.0$ & $\mathbf{100.0 \pm 0.0}$ & $60.0\mathrm{m}$ / $\$0.184$ & $160\mathrm{k}$ in $33.1\mathrm{m}$ (SD $\pm 2.6\mathrm{k}$) \\
\textbf{VeriSegue (Full)} & $\mathbf{163{,}217}$ ($\mathbf{57.84\%}$) & $\mathbf{+20.2\%}$ & $\mathbf{61.2}$ & $\mathbf{62.6}$ & $\mathbf{57.1}$ & $\mathbf{100.0 \pm 0.0}$ & $60.0\mathrm{m}$ / $\$0.164$ & \textbf{MMU $26.4\%$; $160\mathrm{k}$ $15.6\mathrm{m}$ ($2.11\times$)} \\
\midrule
\multicolumn{9}{@{}l@{}}{\textit{(d) Gemmini $16{\times}16$ Systolic Accelerator ($200.0\,\mathrm{min}$ equal horizon, $219{,}285$ RTL; FLAME adapted)}} \\
Local RoCC / bareMetalC & $141{,}359$ ($64.46\%$) & --- & $83.9$ & $78.6$ & $63.3$ & $50.0 \pm 0.0$ & $200\mathrm{m}$ / $\$0$ & CISC $11.2\%$; $1.0/7$ bugs \\
FLAME & $139{,}587$ ($63.66\%$) & $0.0\%$ & $82.3$ & $77.3$ & $62.5$ & $48.1 \pm 4.7$ & $200\mathrm{m}$ / $\$0.174$ & CISC $11.0\%$; $1.2/7$ bugs \\
VeriSegue (No Return) & $159{,}113$ ($72.56\%$) & $+14.0\%$ & $90.8$ & $92.7$ & $71.0$ & $90.3 \pm 2.3$ & $200\mathrm{m}$ / $\$0.040$ & CISC $27.7\%$; $5.2/7$ bugs \\
\textbf{VeriSegue (Full)} & $\mathbf{169{,}049}$ ($\mathbf{77.09\%}$) & $\mathbf{+21.1\%}$ & $\mathbf{90.8}$ & $\mathbf{93.0}$ & $\mathbf{75.8}$ & $\mathbf{91.9 \pm 0.7}$ & $200\mathrm{m}$ / $\mathbf{\$0.039}$ & \textbf{CISC $29.6\%$; $7/7$+D8 ($4.52\times\$$)} \\
\midrule
\multicolumn{9}{@{}l@{}}{\textit{(e) AxCore MPWq4 Block-FP Accelerator ($T=36$ tiles, $2{,}621$ SAT-pruned UB; FLAME adapted)}} \\
Random Baseline & $2{,}615.2$ ($99.78\%$) & --- & $70.4$ & $95.3$ & $97.6$ & $60.8 \pm 1.1$ & $24.1\mathrm{s}$ / $\$0$ & $2.6 \pm 0.6/10$ ($74\%$ miss) \\
FLAME & $2{,}612.6$ ($99.68\%$) & $0.0\%$ & $70.4$ & $94.1$ & $97.5$ & $100.0 \pm 0.0$ & $341\mathrm{s}$ / $\$0.051$ & $6.4 \pm 0.6/10$ ($72$ calls) \\
VeriSegue (No Return) & $2{,}613.8$ ($99.73\%$) & $+18.8\%\text{b}$ & $70.4$ & $94.7$ & $97.6$ & $99.6 \pm 0.9$ & $99.5\mathrm{s}$ / $\$0.024$ & $7.6 \pm 0.6/10$ ($5/5$ D6) \\
\textbf{VeriSegue (Full)} & $\mathbf{2{,}618.0}$ ($\mathbf{99.89\%}$) & $\mathbf{+40.6\%\text{b}}$ & $70.4$ & $\mathbf{97.1}$ & $\mathbf{97.7}$ & $\mathbf{100.0 \pm 0.0}$ & $\mathbf{85.5\mathrm{s}}$ / $\mathbf{\$0.012}$ & \textbf{$9.0{\pm}0.0/10$ ($4.0\times$t, $4.1\times\$$)} \\
\bottomrule
\end{tabular*}
\vspace{2pt}
\parbox{\textwidth}{\scriptsize Rel.\ $\Delta$ denotes relative real-RTL gain vs.\ FLAME (for AxCore, `b' denotes relative defect gain over FLAME as toggle saturates near ceiling). All 80 trajectories across RocketChip, CVA6, SonicBOOM ($60.0\,\mathrm{min}$), and Gemmini ($200.0\,\mathrm{min}$) run past the common horizon and are cut causally at $t \le T_{\max}$. $^\dagger$At FLAME-matched API spend ($\le \$0.200$), RocketChip Full already reaches $71.58\%$ RTL ($69{,}084$) and $99.38\%$ Func in $44.0\,\mathrm{m}$ at $\$0.194$. Full RTL SDs ($n=5$): Rocket $\pm 1{,}370$, CVA6 $\pm 305$, BOOM $\pm 935$, Gemmini $\pm 549$, AxCore $\pm 0.0$.}
\end{table*}

\textbf{Dual-Simulator Parity and Baseline Reproduction.} Original GenHuzz~\cite{genhuzz} trains PPO on VCS condition, line, and FSM bitstrings. We evaluate both on \emph{native VCS-MX (O-2018.09-SP2)} (native Stage-1 training, live CF-UCB scheduling, \texttt{urg} merges, and full-horizon replay of RocketChip and AxCore runs) and on \emph{open-source Verilator} ($15$-run reward audit confirming all Verilator GenHuzz PPO ports stall in $49.9\mathrm{k}$--$50.6\mathrm{k}$ with negative late critic EV of $-0.05$ to $-0.42$).

\begin{figure*}[t]
  \centering
  \includegraphics[width=0.97\textwidth]{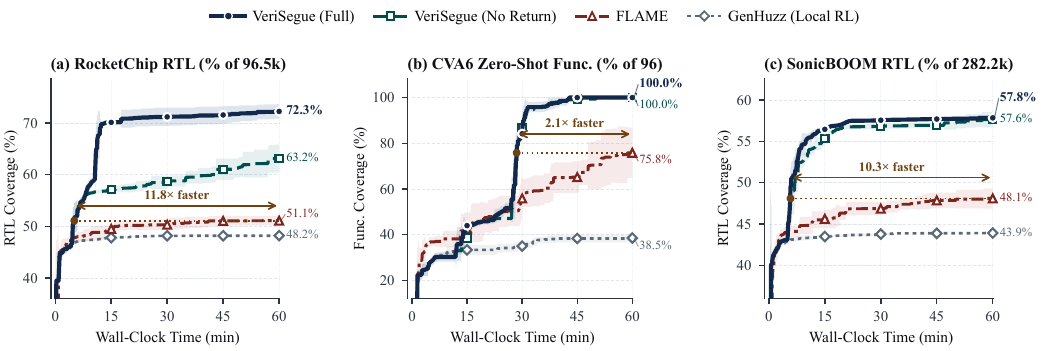}
  \caption{Unified five-seed equal-horizon trajectories across all three RISC-V processors ($0$--$60\,\mathrm{min}$ linear wall-clock axis, shaded $\pm 1\,\mathrm{SD}$, sparse $15\,\mathrm{min}$ checkpoint markers, right-edge $60\,\mathrm{min}$ final means). Dotted ochre bridges mark the time for VeriSegue (Full) to match the best external baseline's $60\,\mathrm{min}$ coverage: (a)~RocketChip real RTL ($11.8\times$ faster); (b)~zero-shot unseen CVA6 functional coverage ($100.0\%$ closure on $5/5$ seeds, $2.1\times$ faster); (c)~superscalar OoO SonicBOOM real RTL ($10.3\times$ faster).}
  \label{fig:cpu_trajectories}
\end{figure*}

\begin{figure*}[t]
  \centering
  \includegraphics[width=\textwidth]{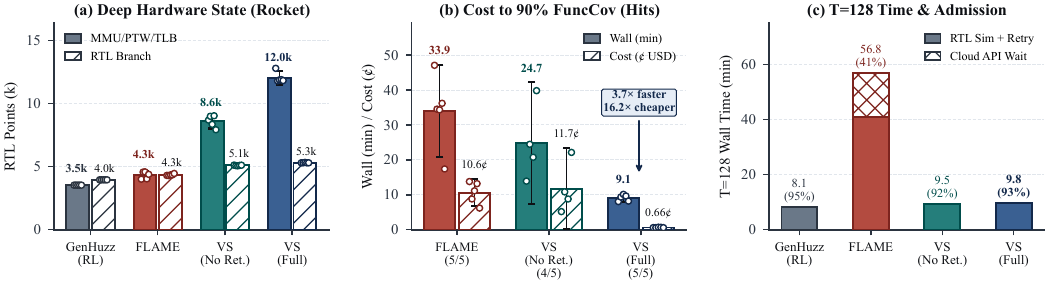}
  \caption{RocketChip deep-state efficacy and resource breakdown ($n=5$, individual seed dots overlaid). (a)~At $60\,\mathrm{min}$, admission-decoupled CF-UCB lifts MMU/PTW/TLB RTL coverage by $2.79\times$ ($12.0\mathrm{k}$ vs.\ $4.3\mathrm{k}$) and Branch by $+22.1\%$; (b)~reaching $90\%$ functional coverage takes $9.1\,\mathrm{min}$ and $0.66\,\text{\textcent}$ ($3.73\times$ faster, $16.2\times$ cheaper than FLAME); (c)~$T=128$ time breakdown shows FLAME loses $34.3\,\mathrm{min}$ to retries ($41\%$ admission vs.\ $93\%$ for VeriSegue).}
  \label{fig:rocket_subsystems}
\end{figure*}

To provide an authoritative comparison against FLAME~\cite{flame}, we cloned the official repository (\texttt{superpung/flame}) and preserved its constrained-random (CRV) cold start, ISA/RTL knowledge-base RAG retrieval, three-turn CoT pipeline, and positive/error feedback loop under identical harnesses and oracles. Because native GenHuzz and FLAME target RISC-V processors and do not support AI accelerators, we additionally ported FLAME's RAG+CoT loop onto our Gemmini and AxCore adapters as a strong adapted accelerator baseline. Across all five targets, we run four matched arms over five seeds $\{613, 719, 823, 947, 1061\}$ ($100$ runs total) with exact prompt, token, and USD cost logging ($T=0.2$): (1)~\emph{Local RL / Baseline} (GenHuzz, Local RoCC, or Random); (2)~\emph{FLAME} (\texttt{DeepSeek-V4.1-Flash}); (3)~\emph{VeriSegue (No Deep-State Return)}; and (4)~\emph{VeriSegue (Full Bidirectional)} (evaluating frozen \texttt{DeepSeek-V4.1-Flash}, \texttt{Qwen3.8-Max}, and \texttt{GLM-5.2}/\texttt{5.3}). Every run is audited under equal simulation count, strictly matched equal wall-clock horizons ($60.0\,\mathrm{min}$ across all three RISC-V processors and $200.0\,\mathrm{min}$ on Gemmini), equal API cost (including RocketChip's FLAME-matched $\le \$0.200$ checkpoint), and equal target coverage using deterministic $4$-worker FIFO scheduling over uncached simulation and API latencies. Stage-1 online RL training, BF16 VRAM profiling (Table~\ref{tab:vcs_native}(a), Fig.~\ref{fig:stage1eff}), and optional Path-B $25\,\mathrm{s}$ KL-replay (Table~\ref{tab:vcs_native}(c), Table~\ref{tab:deepstate_attr}) run on a local workstation GPU; in multi-seed equal-horizon evaluation (Table~\ref{tab:e2e_eval}), Stage~1 deploys the once-pretrained frozen policy under Path~A ($0\,\text{GPU-h}$ retraining, ${\sim}20\,\mathrm{ms}$/test forward latency vs.\ multi-second RTL simulation), while Stage~2 queries frozen cloud LLM APIs and executes three-oracle RTL verification on local CPUs. Multi-vendor counterfactuals further span eight LLMs across five vendors: OpenAI (GPT-6.1-Sol, GPT-6.2-Astra, GPT-6.2-Sol, GPT-6-Astra), Alibaba (Qwen3.8-Max), DeepSeek (DeepSeek-V4.1-Flash), Zhipu (GLM-5.3), and Moonshot (Kimi-K3.1-Pro). We report $\mathrm{Mean}\pm\mathrm{SD}$, $95\%$ $t$-CIs, exact two-sided Mann--Whitney $MW_p$ (minimum $0.0079$ for $n=5$), Cliff's $\delta$, and Hedges' $g$, strictly separating single-run statistics from multi-run unions.

\subsection{VCS Five-Metric Validation}
\label{sec:vcs_eval}

Table~\ref{tab:vcs_native} and Fig.~\ref{fig:vcs_panorama}(a) validate VeriSegue directly inside VCS-MX on RocketTile. In native Stage-1 training (Table~\ref{tab:vcs_native}(a)), unmasked GenHuzz PPO fails compilation ($0\%$) and masked PPO exhibits negative late critic EV ($-0.085$ to $-0.185$), whereas VeriSegue GRPO exceeds masked PPO on all five VCS metrics in $1.96\times$ less wall time ($19.6$ vs.\ $38.5\,\mathrm{min}$). Live online CF-UCB scheduling (Table~\ref{tab:vcs_native}(b)) outperforms uniform random selection across every metric ($+8.95\,$pp FSM) while cutting variance by $20\times$--$66\times$, and native \texttt{urg} VDB merges (Table~\ref{tab:vcs_native}(c)) confirm that closed-loop $\mathrm{S1}\to\mathrm{S2}\to\mathrm{S1}$ raises FSM coverage by $+13.59\,$pp over Stage~1 alone. Replaying all four $60\,\mathrm{min}$ equal-horizon arms ($1{,}688$ unique RV64GC programs, Table~\ref{tab:vcs_native}(d)) yields unanimous $5/5$ wins for VeriSegue (Full) over FLAME across all six VCS metrics---lifting Score by $+18.9\%$ ($+9.17\,$pp, $p=0.0079$; $63.29\%$ \texttt{mod\_def}), Toggle by $+37.1\%$, and FSM by $+32.1\%$ ($p=0.0238$; $36.58\%$ \texttt{mod\_def})---while also exceeding No-Return on every metric ($5/5$ wins). VCS additionally confirms bit-exact parity for \texttt{CPU-BUG-05} ($0.8\,\mathrm{s}$/case) and AxCore ($0$ mismatches across $25$ replayed runs; $97.67\%$ core Toggle, $0/0$ FSM).

\subsection{Five-Seed Equal-Horizon End-to-End Evaluation}
\label{sec:e2e_eval}

Table~\ref{tab:e2e_eval} and Figs.~\ref{fig:cpu_trajectories}--\ref{fig:rocket_subsystems} summarize the five-seed equal-horizon study against FLAME. On RocketChip (Table~\ref{tab:e2e_eval}(a), Fig.~\ref{fig:cpu_trajectories}(a)), VeriSegue (Full) already hits $58.86\%$ real RTL and $90.83\%$ functional coverage at the $T=128$ equal-simulation checkpoint in $9.84\,\mathrm{min}$ and $\$0.0061$ ($93.1\%$ admission), whereas FLAME spends $56.8\,\mathrm{min}$ ($34.3\,\mathrm{min}$ on $53$ retries at $40.8\%$ admission, Fig.~\ref{fig:rocket_subsystems}(c)). Over the matched $60.0\,\mathrm{min}$ horizon, FLAME saturates at $51.11\%$ real RTL ($49{,}330$ points, $40.1\%$ admission), whereas closed-loop \texttt{SynLib} synthesis and multi-cycle $\mathrm{S2}\to\mathrm{S1}$ stacking push VeriSegue (Full) to $72.27\%$ real RTL ($+41.4\%$ over FLAME, Hedges' $g=+15.86$, $4/5$ seeds hitting $100\%$ functional coverage)---crossing FLAME's $60\,\mathrm{min}$ RTL mark in $5.1\,\mathrm{min}$ ($11.8\times$ faster). Even under strict FLAME-matched API spend ($\le \$0.200$), VeriSegue (Full) already reaches $71.58\%$ RTL and $99.38\%$ functional coverage in $44.0\,\mathrm{min}$ at $\$0.194$.

Admission-decoupled CF-UCB improves every structural dimension simultaneously, raising MMU/PTW/TLB RTL coverage by $2.79\times$ and RTL branches by $+22.1\%$ (Fig.~\ref{fig:rocket_subsystems}(a)). Comparing Full against No-Return at $60.0\,\mathrm{min}$ isolates bidirectional return cleanly: deep-state return adds $+8{,}804$ real RTL ($5/5$ wins, $p=0.0079$), lifts $65\mathrm{k}$-point reliability from $0/5$ to $5/5$ seeds, and reaches $90\%$ functional coverage $2.72\times$ faster while spending $14.1\%$ less total API budget ($\$0.286$ vs.\ $\$0.333$, Fig.~\ref{fig:rocket_subsystems}(b)).

\subsection{Stage-1 RL Isolation}
\label{sec:s1_eval}

\begin{figure}[t]
  \centering
  \includegraphics[width=0.88\columnwidth]{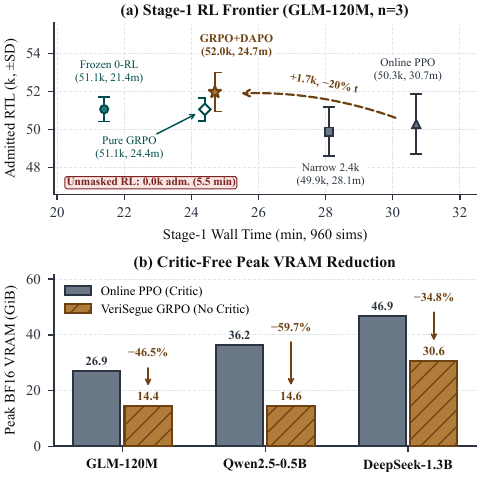}
  \caption{Controlled $960$-simulation Stage-1 RL isolation on RocketChip (ISA-GLM-120M, fixed vocabulary, mask, and reward, $n=3$). (a)~Critic-free GRPO+DAPO exceeds masked PPO by $+1{,}684$ points in $19.5\%$ less time while unmasked RL collapses to $0.0\mathrm{k}$; (b)~dropping the critic cuts peak BF16 VRAM by $34.8\%$--$59.7\%$ across $120\mathrm{M}$--$1.3\mathrm{B}$ backbones.}
  \label{fig:stage1eff}
\end{figure}

Controlled $960$-simulation ablations on RocketChip (Fig.~\ref{fig:stage1eff}, identical $120\mathrm{M}$ backbone, $11{,}642$ vocabulary, syntax mask, and reward, $n=3$) strictly isolate PPO from GRPO. Grammar masking lifts compilation from $0\%$ to $100\%$. Replacing masked PPO ($50{,}297 \pm 1{,}564$ points, $30.7\,\mathrm{min}$, late critic $\mathrm{EV}=-0.025$) with critic-free GRPO reduces SD by $2.55\times$ in $20.5\%$ less wall time, and asymmetric DAPO clipping ($\epsilon_{\mathrm{high}}=0.28$) reaches $51{,}981 \pm 1{,}025$ points ($+1{,}684$ over PPO in $19.5\%$ less time, exceeding frozen $0$-RL $51.1\mathrm{k}$, offline RFT $51.7\mathrm{k}$, and DPO $51.2\mathrm{k}$) while slashing peak BF16 VRAM by $34.8\%$--$59.7\%$ across backbones (Fig.~\ref{fig:stage1eff}(b)). Forking all eight frontier LLMs from the shared $W=20$ handoff ($\tau^* = 50{,}817$) confirms that extending local GRPO adds only $+145$ points, whereas fixed-menu CF-UCB maintains $100\%$ admission ($60{,}888 \pm 132$ at $R=28$, beating DIFUZZRTL $48.3\mathrm{k}$, $13.3\,\mathrm{h}$ Verilator GenHuzz $52.6\mathrm{k}$, and Cascade-Priv $55.2\mathrm{k}$), and dynamic \texttt{SynLib} synthesis breaks the narrow fixed-menu ceiling to reach $69.8\mathrm{k}$.

\subsection{Handoff, Slicing Depth, Counterfactuals, and Component Ablations}
\label{sec:ablation_cost}

\begin{figure}[t]
  \centering
  \includegraphics[width=0.92\columnwidth]{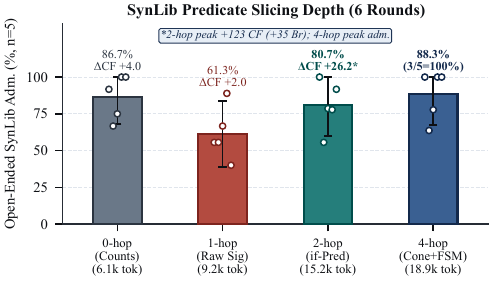}
  \caption{Backward RTL slicing depth on open-ended \texttt{SynLib} synthesis (\texttt{DeepSeek-V4.1-Flash}, $6$ rounds, $n=5$). $2$-hop Verilog \texttt{if}-predicate slicing peaks control-flow expansion ($\Delta\mathrm{CF}=+26.2$, peak $+123$), while $4$-hop cones with FSM arcs achieve the highest legal admission ($88.3\%$, $3/5$ runs $100\%$).}
  \label{fig:handoff_hops}
\end{figure}

\textbf{Adaptive Handoff and Slicing Depth (Fig.~\ref{fig:handoff_hops}).} Across six handoff policies at $T=128$ (\texttt{DeepSeek-V4.1-Flash}, $n=5$), Local-Only ($100/0$) stalls at $42.71\%$ functional coverage ($45.9\mathrm{k}$ RTL) and Cloud-Only ($0/100$) spends $\$0.0178$ yet trails at $85.63\%$ ($55.8\mathrm{k}$ RTL). Adaptive $\gamma_{\mathrm{cf}}$ triggers organically in $5/5$ seeds at $\mathrm{sim}=40$, reaching peak functional ($91.25 \pm 1.19\%$) and RTL ($57{,}323 \pm 151$) coverage at $2.4\times$ lower API cost ($\$0.0074$) than Cloud-Only without hand-tuned splits ($75/25$: $86.0\%$; $50/50$: $89.6\%$; $25/75$: $89.8\%$). Ablating backward slicing depth on open-ended \texttt{SynLib} synthesis (\texttt{DeepSeek-V4.1-Flash}, $6$ rounds, Fig.~\ref{fig:handoff_hops}) shows that raw $1$-hop signal names drop admission to $61.3\%$, whereas $2$-hop Verilog \texttt{if}-predicates ($15.2\mathrm{k}$ tokens) maximize control-flow growth ($\Delta\mathrm{CF}=+26.2$, peak $+123$) and $4$-hop cones with driving CSRs and FSM arcs ($18.9\mathrm{k}$ tokens) achieve the highest legal admission ($88.3\%$, $3/5$ seeds at $100\%$).

\begin{figure*}[t]
  \centering
  \includegraphics[width=\textwidth]{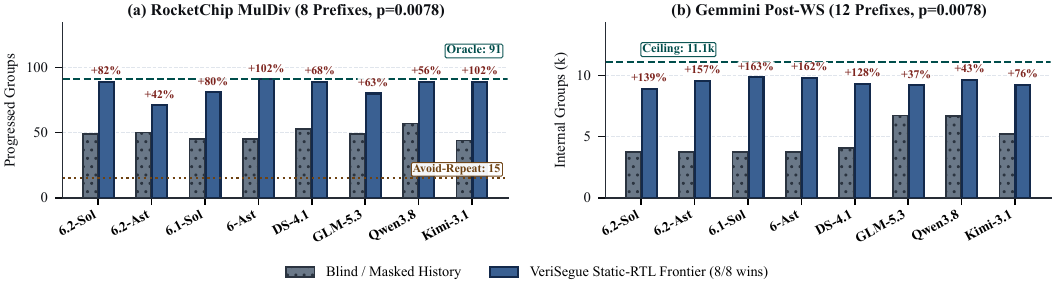}
  \caption{Same-last-action counterfactuals across eight LLMs (five vendors, two-sided Wilcoxon $p=0.0078$). Residual RTL dossiers lift selection accuracy on $8/8$ models across (a)~RocketChip MulDiv ($+73.2\%$, approaching the $91$-group oracle) and (b)~Gemmini post-weight-stationary schedules ($+100.4\%$).}
  \label{fig:counterfactuals}
\end{figure*}

\textbf{Counterfactuals, Component Ablations, and Cost.} Holding model, state, menu, and history identical across all eight LLMs (Fig.~\ref{fig:counterfactuals}), residual RTL dossiers lift selection accuracy on $8/8$ models ($p=0.0078$) across RocketChip MulDiv ($+73.2\%$, $84.9/91$ oracle hits; Yosys SAT certifies all $3$ remaining branches UNSAT) and Gemmini post-WS schedules ($+100.4\%$ internal groups, $+60.7\%$ RTL points, $p=0.00098$). Table~\ref{tab:ablation_cost} confirms monotonic contributions from every component: omitting Stage~1 loses $13{,}549$ points, dropping constructors cuts admission to $60\%$, and full RTL CF-UCB ($58{,}855 \pm 189$) outperforms unranked prompting, blind constructors, and coarse counts ($p=0.00781$), while disabling $\mathcal{G}_{\mathrm{cov}}$ in~(\ref{eq:gate}) admits $9/9$ faulty tests ($+2{,}130$ spurious points). Amortized Stage-1 pretraining takes $1.85\,\text{GPU-h}$ once on RocketChip ($0\,\text{GPU-h}$ on zero-shot CVA6), while per-run equal-horizon API spend spans $\$0.012$ (AxCore) and $\$0.039$--$\$0.164$ (Gemmini, CVA6, SonicBOOM) to $\$0.194$--$\$0.286$ (RocketChip).

\begin{table*}[t]
\centering
\caption{Two-regime necessity and Stage-2 CF-UCB component ablations on RocketChip ($R=8$ fixed menu, $n=3$; S1=Stage~1, Ct=Constructors, Cnt=Counts, RTL=Backward slice, UCB/Tb=CF-UCB+Tabu). Every module contributes monotonically to admitted coverage.}
\label{tab:ablation_cost}
\footnotesize
\setlength{\tabcolsep}{0pt}
\renewcommand{\arraystretch}{1.04}
\begin{tabular*}{\textwidth}{@{\extracolsep{\fill}}lcccccrrr@{}}
\toprule
\textbf{Ablation Variant} & \textbf{S1} & \textbf{Ct} & \textbf{Cnt} & \textbf{RTL} & \textbf{UCB/Tb} & \textbf{Adm. (\%)} & \textbf{Final (Mean$\pm$SD)} & \textbf{S2 $\Delta$} \\
\midrule
Stage-1 Handoff ($\tau^*$, no Stage~2) & \checkmark & --- & --- & --- & --- & $100$ & $50{,}817 \pm 0$ & $0$ \\
Stage-2 From Scratch ($\mathcal{C}_0=\emptyset$) & --- & \checkmark & \checkmark & \checkmark & \checkmark & $100$ & $45{,}307 \pm 492$ & $-5{,}511$ \\
Raw-ASM (No Constructors) & \checkmark & --- & \checkmark & \checkmark & \checkmark & $60$ & $53{,}622 \pm 1{,}464$ & $+2{,}804$ \\
Locked Minimal Params ($\theta_{\min}$) & \checkmark & \checkmark & --- & --- & --- & $100$ & $54{,}892 \pm 0$ & $+4{,}075$ \\
Flat Unranked RTL (No CF-UCB/Tabu) & \checkmark & \checkmark & \checkmark & \checkmark & --- & $100$ & $56{,}953 \pm 684$ & $+6{,}136$ \\
Blind Constructors (No Frontier) & \checkmark & \checkmark & --- & --- & Tb & $100$ & $57{,}663 \pm 518$ & $+6{,}846$ \\
Coarse Counts + CF-UCB & \checkmark & \checkmark & \checkmark & --- & \checkmark & $100$ & $58{,}414 \pm 395$ & $+7{,}597$ \\
\textbf{Full VeriSegue CF-UCB (RTL Slice)} & \checkmark & \checkmark & \checkmark & \checkmark & \checkmark & $\mathbf{100}$ & $\mathbf{58{,}855 \pm 189}$ & $\mathbf{+8{,}038}$ \\
\bottomrule
\end{tabular*}
\end{table*}

\section{Cross-Architecture Generalization, Zero-Shot Transfer, and Deep-State Return}
\label{sec:generalization}

\subsection{Zero-Shot CVA6, OoO SonicBOOM, and AI Accelerators}

\textbf{Zero-Shot Transfer to Unseen SystemVerilog CVA6 and OoO SonicBOOM (Table~\ref{tab:e2e_eval}(b)--(c), Fig.~\ref{fig:cpu_trajectories}(b)--(c), Table~\ref{tab:target_and_tracesmt}(a)).} Freezing Stage-1 RocketChip weights ($0\,\text{GPU-h}$ retraining), the generic RV64GC prompt, and CF-UCB parameters with zero core edits, VeriSegue (Full) transfers zero-shot to unseen SystemVerilog CVA6 over the matched $60.0\,\mathrm{min}$ horizon, winning $5/5$ seeds ($MW_p=0.0079$) over both FLAME ($+24.1\%$ real RTL, $+24.2\,$pp functional coverage) and No-Return ($+3{,}290$ real RTL, $5/5$ wins) while achieving $100.0\%$ total and MMU functional coverage across all five seeds (vs.\ $75.8\%$ total and $27.8\%$ MMU for FLAME) at $1.18\times$ lower API cost ($\$0.124$ vs.\ $\$0.146$), crossing $65\mathrm{k}$ RTL on $5/5$ seeds in $38.1\,\mathrm{min}$ vs.\ $0/5$ for all other arms; differential Spike triage simultaneously exposes \texttt{CVA6-D1} across $5/5$ seeds (vs.\ $2/5$ for FLAME; Section~\ref{sec:forensic_ledger}). On superscalar OoO SonicBOOM at $60.0\,\mathrm{min}$, VeriSegue (Full) wins $5/5$ seeds over FLAME across every metric ($57.84\%$ RTL, $100.0\%$ functional coverage on $5/5$ seeds vs.\ $86.0\%$), crosses $150\mathrm{k}$ RTL on $5/5$ seeds in $8.1\,\mathrm{min}$ ($0.38\,\text{\textcent}$, Table~\ref{tab:target_and_tracesmt}(a)) vs.\ $0/5$ for FLAME, reaches $160\mathrm{k}$ $2.11\times$ faster than No-Return ($15.6$ vs.\ $33.1\,\mathrm{min}$) with $2.77\times$ lower SD and $11.1\%$ lower API cost, and exceeds $2{,}500$-case Cascade-Priv ($155.6\mathrm{k}$ in $25{,}700\,\mathrm{s}$) in $185.5\,\mathrm{s}$. Cross-vendor scaling over all eight LLMs ($24$ runs/DUT, $100\%$ admission) further confirms model-agnosticity, lifting SonicBOOM MMU coverage by $+104.3\%$ mean ($+107.1\%$ best), Gemmini by $+68.8\%$ over bareMetalC, and AxCore joint D1\&D2 exposure to $79.2\%$ mean ($100\%$ best).

\begin{figure*}[t]
  \centering
  \includegraphics[width=0.97\textwidth]{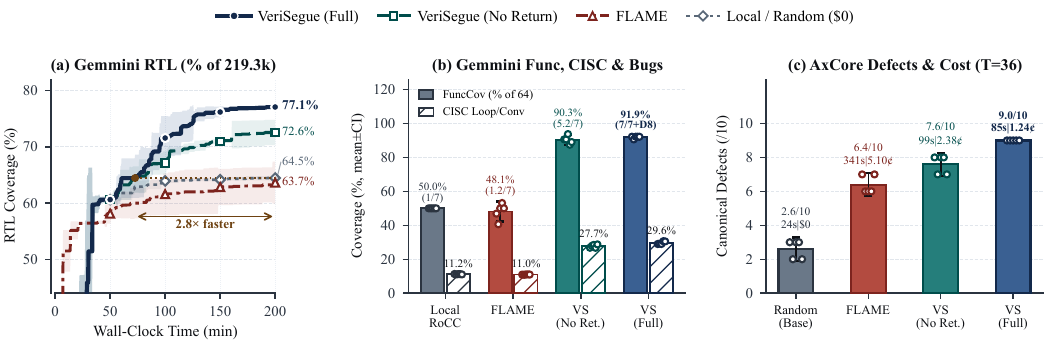}
  \caption{AI accelerator evaluation across Gemmini ($0$--$200\,\mathrm{min}$ equal horizon) and AxCore ($T=36$ tiles, $n=5$; FLAME adapted from its CPU-only original flow via our modular harness). (a)~Gemmini real RTL ($\%$ of $219.3\mathrm{k}$, shaded $\pm 1\,\mathrm{SD}$, right-edge $200\,\mathrm{min}$ means): VeriSegue (Full) reaches $77.09\%$ ($+21.1\%$ vs.\ FLAME) and crosses the best external baseline's $200\,\mathrm{min}$ mark $2.8\times$ faster (dotted ochre bridge); (b)~Gemmini $200\,\mathrm{min}$ functional coverage ($91.88\%$), CISC \texttt{LoopMatmul}/\texttt{LoopConv} FSM ($29.64\%$, $2.70\times$ FLAME), and $7.0/7$ defects plus live 0-day \texttt{GEM-D8}; (c)~AxCore verified defects ($9.00 \pm 0.00/10$ in $85.5\,\mathrm{s}$ and $1.24\,\text{\textcent}$, $4.0\times$ faster and $4.1\times$ cheaper than FLAME).}
  \label{fig:accel_eval}
\end{figure*}

\textbf{Unified Support for AI Accelerators: Gemmini and AxCore (Table~\ref{tab:e2e_eval}(d)--(e), Fig.~\ref{fig:accel_eval}).} A distinctive architectural advantage of VeriSegue is unifying RISC-V processors and heterogeneous AI accelerators behind the same four-hook adapter $\mathcal{A}_d$: prior local-RL and LLM verifiers such as GenHuzz~\cite{genhuzz} and FLAME~\cite{flame} target processor instruction streams and lack native support for C+RoCC systolic arrays or approximate block-FP engines. Even when compared against our adapted FLAME accelerator baseline over strictly matched horizons, VeriSegue (Full) dominates both targets cleanly. On Gemmini at $200.0\,\mathrm{min}$ (Fig.~\ref{fig:accel_eval}(a)--(b)), extending Local RoCC and FLAME to full $200\,\mathrm{min}$ saturates both near $63.7\%$--$64.5\%$ RTL and $11.0\%$--$11.2\%$ CISC FSM ($1.0$--$1.2/7$ defects), whereas VeriSegue (Full) crosses the best external baseline's $200\,\mathrm{min}$ mark $2.8\times$ faster ($72.7\,\mathrm{min}$), climbs to $77.09\%$ RTL ($+21.1\%$ vs.\ FLAME; $+4.53\,$pp over No-Return, $5/5$ wins) and $2.70\times$ higher CISC \texttt{LoopMatmul}/\texttt{LoopConv} FSM coverage ($29.64\%$), and sweeps all $7.0/7$ canonical defects plus $5/5$ autonomous exposure of live upstream 0-day \texttt{GEM-D8} at $4.52\times$ lower API cost ($\$0.039$ vs.\ $\$0.174$). On AxCore (Fig.~\ref{fig:accel_eval}(c)), random tiles saturate structural toggle ($99.78\%$, $97.67\%$ on VCS) yet miss $74\%$ of canonical RTL defects ($2.6/10$); VeriSegue (Full) reaches the exact $99.89\%$ FP4 structural ceiling, hits $100\%$ functional coverage ($3.2\times$ faster than FLAME), and exposes $9.00 \pm 0.00/10$ defects with zero variance in $85.5\,\mathrm{s}$ and $\$0.0124$ ($4.0\times$ faster and $4.1\times$ cheaper than FLAME's $6.4/10$). Pairing dynamic Stage-2 closure ($9/10$) with Tier-1 TraceSMT/SAT on \texttt{AX-D7} achieves complete $10/10$ defect closure.

\begin{table*}[t]
\centering
\caption{(a)~Equal-target resource savings vs.\ FLAME within the $60.0\,\mathrm{min}$ equal horizon across all three RISC-V processors ($n=5$); and (b)~Hybrid two-tier TraceSMT solver hierarchy matching local SAT/algebraic solvers on shallow datapaths with VPI-guided LLM inversion on $100\mathrm{k}$-SLOC sequential cores.}
\label{tab:target_and_tracesmt}
\footnotesize
\setlength{\tabcolsep}{0pt}
\renewcommand{\arraystretch}{1.04}
\parbox{\textwidth}{\textit{(a) Equal-Target Resource Savings Within $60.0\,\mathrm{min}$ ($n=5$)}}\\[2pt]
\begin{tabular*}{\textwidth}{@{\extracolsep{\fill}}l cccl@{}}
\toprule
\textbf{Coverage Target} & \textbf{FLAME} & \textbf{VS (No Ret.)} & \textbf{VeriSegue (Full)} & \textbf{Speedup / Takeaway} \\
\midrule
Rocket $\mathrm{Func}\ge 90\%$ & $5/5, 33.9\mathrm{m}, 10.6\text{\textcent}$ & $4/5, 24.7\mathrm{m}, 11.8\text{\textcent}$ & $\mathbf{5/5, 9.1\mathrm{m}, 0.66\text{\textcent}}$ & $\mathbf{3.73\times\mathrm{t},\; 16.2\times\$}$ \\
Rocket $\mathrm{RTL}\ge 60\mathrm{k}$ & $0/5$ & $3/5, 47.3\mathrm{m}, 25.7\text{\textcent}$ & $\mathbf{5/5, 11.2\mathrm{m}, 0.95\text{\textcent}}$ & $\mathbf{5/5}$ ($5/5{\ge}65\mathrm{k}$ in $11.3\mathrm{m}$) \\
CVA6 $\mathrm{Func}\ge 85\%$ & $2/5, 49.7\mathrm{m}, 11.6\text{\textcent}$ & $5/5, 30.5\mathrm{m}, 0.66\text{\textcent}$ & $\mathbf{5/5, 30.6\mathrm{m}, 0.66\text{\textcent}}$ & $\mathbf{5/5}$ vs.\ $2/5$ ($5/5{\ge}65\mathrm{k}$ in $38.1\mathrm{m}$) \\
BOOM $\mathrm{RTL}\ge 150\mathrm{k}$ & $0/5$ & $5/5, 9.5\mathrm{m}, 0.42\text{\textcent}$ & $\mathbf{5/5, 8.1\mathrm{m}, 0.38\text{\textcent}}$ & $\mathbf{5/5}$ ($160\mathrm{k}$ $2.11\times$ faster: $15.6\mathrm{m}$) \\
\bottomrule
\end{tabular*}

\vspace{4pt}
\parbox{\textwidth}{\textit{(b) Hybrid Two-Tier TraceSMT Solver Hierarchy Across Hardware Depth}}\\[2pt]
\begin{tabular*}{\textwidth}{@{\extracolsep{\fill}}l ccccl@{}}
\toprule
\textbf{Hardware Regime} & \textbf{Yosys SAT} & \textbf{Alg.\ Secant} & \textbf{Non-LLM VPI/Rnd} & \textbf{TraceSMT-LLM} & \textbf{Tier Assignment} \\
\midrule
Tier~1: Shallow $1$-DFF (AxCore G1--G5) & $3/3$ ($2.5$--$3.3\mathrm{s}$, $\$0$) & $\mathbf{5/5}$ ($1$--$3$ ev., $\$0$) & $40\%$--$80\%$ (G5 $0/5$) & $14/15$--$15/15$ & Tier-1 SAT/Secant ($\$0$) \\
Tier~2: Deep Seq.\ (Rocket $100\mathrm{k}$L, B1--B4) & Fails ($0.47\mathrm{s}$) & N/A (multi-inst) & $\mathbf{0/20}$ ($0/5$ all 4) & $\mathbf{3/4\text{--}4/4}$ ($+1.8\mathrm{k}$) & Tier-2 VPI TraceSMT-LLM \\
\bottomrule
\end{tabular*}
\end{table*}

\subsection{Frontier Mechanisms, TraceSMT Hierarchy, and Deep-State Attribution}
\label{sec:deepstate_eval}

\textbf{Dynamic Constructor Synthesis (\texttt{SynLib}) and Boundary Batching.} Synthesizing parameterized constructors breaks the fixed-menu ceiling ($61.6\mathrm{k}$), lifting single-run RocketChip coverage to $69.8\mathrm{k}$ (multi-run unions $>71\mathrm{k}$ on RocketChip and $>171\mathrm{k}$ on SonicBOOM). All $17/17$ promoted RocketChip constructors transfer zero-shot to SonicBOOM (up to $+7{,}894$ points, exposing \texttt{CPU-BUG-03} and \texttt{CPU-BUG-06}). Comparing $K=5$ boundary batching against combinatorial $2$-wise/$3$-wise sweeps ($n=5$, $444$ Gemmini and $450$ AxCore runs) shows that while 1D edge sweeps reach \texttt{GEM-D2} ($4/5$), pairwise, $3$-wise, and random sweeps all score $0/5$ on coupled defects \texttt{GEM-D1}, \texttt{GEM-SW-D7}, \texttt{GEM-D8}, and \texttt{AX-D7}, where semantic pairing and TraceSMT are decisive.

\textbf{Hybrid Two-Tier TraceSMT Hierarchy (Table~\ref{tab:target_and_tracesmt}(b)).} Matching solver complexity to hardware depth yields a clean two-tier flow under identical internal telemetry. In Tier~1 (shallow $1$-DFF datapaths, AxCore \texttt{G1}--\texttt{G5}), local Yosys SAT (\texttt{-seq 2}) solves \texttt{G1}--\texttt{G3} (including \texttt{AX-D7}) in $2.5$--$3.3\,\mathrm{s}$ at $\$0$ and algebraic secant solving resolves $5/5$ targets in $1$--$3$ evaluations ($0.01$--$0.04\,\mathrm{s}$). In Tier~2 ($100\mathrm{k}$-SLOC sequential RocketChip branches \texttt{B1}--\texttt{B4}, $141$ unique runs), direct SAT fails ($0.47\,\mathrm{s}$) and non-LLM VPI mutation given the exact same signal telemetry scores $0/20$, whereas TraceSMT-LLM (\texttt{DeepSeek-V4.1-Flash} and \texttt{Qwen3.8-Max}) closes $3/4$--$4/4$ branches ($+1.2\mathrm{k}$--$1.8\mathrm{k}$ real RTL points) by synthesizing legal privileged instruction sequences.

\textbf{Dual-Simulator Deep-State Attribution (Table~\ref{tab:deepstate_attr}).} Table~\ref{tab:deepstate_attr} isolates deep-state context transfer from policy adaptation on RocketChip under Verilator ($960$ sims) and VCS ($40$ deep-state programs/seed). Frozen context transfer (Path~A) provides ${\sim}85\%$--$90\%$ of the return gain at zero retraining: prepending $\mathcal{S}_{\mathrm{seed}}$ lifts frozen Stage~1 by $+11{,}417$ architectural real points ($+25.8\%$) and $65\times$ more novel points beyond the $66.7\mathrm{k}$ union, while $40$ frozen deep-state programs beat $960$ cold-reset Stage-1 tests across all five VCS metrics (FSM $19.44\% \to 26.85\%$, outperforming seed mutation at $70\%$ vs.\ $35\%$ admission). Adding $25\,\mathrm{s}$ of KL-anchored replay (Path~B) avoids un-anchored entropy collapse ($59.4\mathrm{k} < 60.5\mathrm{k}$) and pushes coverage to $64{,}222$ on Verilator and $29.32\%$ VCS FSM ($+9.88\,$pp; multi-run unions reach $68.5\mathrm{k}$ on RocketChip and $176.4\mathrm{k}$ on SonicBOOM).

\section{Defect Discovery, Hardware Patch, and AI Impact}
\label{sec:defects_ai}

Structural coverage maps RTL state space yet cannot guarantee functional correctness, as demonstrated on AxCore in Fig.~\ref{fig:vcs_panorama}(b) and Fig.~\ref{fig:accel_eval}(c). Below we audit Gemmini reference--RTL divergences, AxCore bit-level faults and our formally checked $45\,\mathrm{nm}$ patch, expanded ImageNet-1K ($n=1{,}000$) and quantized-LLM workload attribution, RISC-V specification compliance proofs (\texttt{CPU-BUG-05} and live upstream 0-day \texttt{GEM-D8}, \texttt{ucb-bar/libgemmini} Issue~\#10), and the $20$-root-cause forensic ledger.

\subsection{Gemmini Spike--RTL Divergences and Oracle Escapes}
\begin{table*}[t]
\centering
\caption{Dual-simulator cold-reset vs.\ deep-state attribution on RocketChip ($n=3$): Verilator $960$-simulation architectural real mask ($92{,}553$ points) vs.\ VCS-MX ($40$ deep-state programs/seed vs.\ $960$-test Cold-Reset Stage~1). Deep-state context transfer accounts for ${\sim}85\%$--$90\%$ of the return gain at zero retraining.}
\label{tab:deepstate_attr}
\footnotesize
\setlength{\tabcolsep}{1.6pt}
\renewcommand{\arraystretch}{1.04}
\begin{tabular*}{\textwidth}{@{\extracolsep{\fill}}l rrr rrrrr@{}}
\toprule
 & \multicolumn{3}{c}{\textbf{Verilator ($960$ Sims, $92{,}553$ Arch.\ Mask)}} & \multicolumn{5}{c}{\textbf{VCS (5 Metrics, \%)}} \\
\cmidrule(lr){2-4} \cmidrule(lr){5-9}
\textbf{Initialization \& Adaptation Arm} & \textbf{Total (Mean$\pm$SD)} & \textbf{Arch.\ Real (\%)} & \textbf{New $>$ Un.} & \textbf{Line} & \textbf{Cond} & \textbf{FSM (Hit)} & \textbf{Branch} & \textbf{Toggle} \\
\midrule
Cold-Reset + Frozen & $51{,}076 \pm 644$ & $44{,}272$ ($47.8$) & $+35$ & --- & --- & --- & --- & --- \\
Cold-Reset + GRPO & $51{,}981 \pm 1{,}025$ & $45{,}166$ ($48.8$) & $+314$ & $56.62$ & $73.44$ & $19.44$ ($21.0$) & $59.62$ & $27.15$ \\
Prefixes-Only & $41{,}913 \pm 0$ & $36{,}206$ ($39.1$) & $+36$ & $52.46$ & $67.09$ & $6.48$ ($7.0$) & $55.79$ & $19.08$ \\
Deep-State + Seed-Mutation & $61{,}984 \pm 724$ & $54{,}798$ ($59.2$) & $+1{,}269$ & $58.62$ & $76.01$ & $22.84$ ($24.7$) & $61.78$ & $28.34$ \\
Deep-State + Frozen & $62{,}835 \pm 1{,}258$ & $55{,}688$ ($60.2$) & $+2{,}268$ & $58.77$ & $76.08$ & $26.85$ ($29.0$) & $61.89$ & $31.01$ \\
\textbf{Deep-State + KL-Replay} & $\mathbf{64{,}222 \pm 621}$ & $\mathbf{57{,}010}$ ($\mathbf{61.6}$) & $\mathbf{+3{,}235}$ & $\mathbf{59.58}$ & $\mathbf{78.33}$ & $\mathbf{29.32}$ ($31.7$) & $\mathbf{63.22}$ & $\mathbf{31.44}$ \\
\bottomrule
\end{tabular*}
\end{table*}

\begin{figure*}[t]
  \centering
  \includegraphics[width=0.82\textwidth]{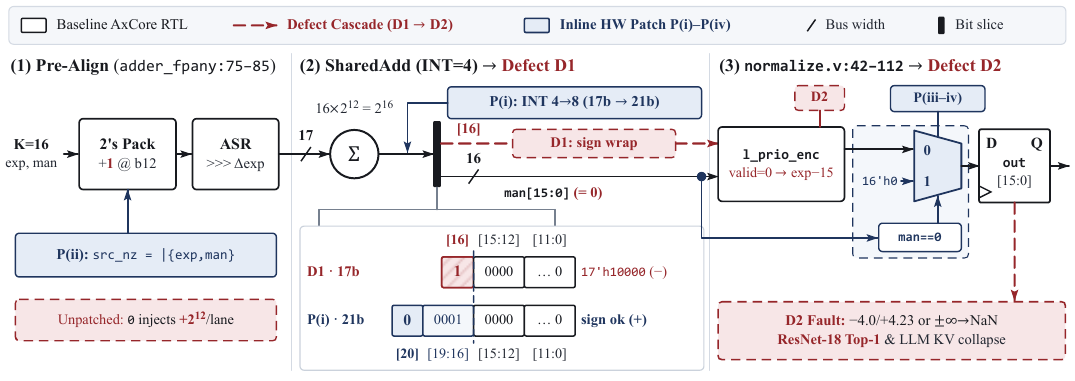}
  \caption{Bit-level RTL cascade of shipped AxCore defects D1 and D2 with inline four-part hardware patches P(i)--P(iv). (1)~Pre-Align injects an unconditional implicit-$1$ at bit~$12$ ($+2^{12}$/lane), gated by \textbf{P(ii)}. (2)~SharedAdd (\texttt{INT=4}) wraps $17$-bit \texttt{man\_sum} to \texttt{17'h10000} (\textbf{D1}, sign bit \texttt{[16]} flipped, \texttt{man[15:0]=0}); widening \texttt{INT} $4{\to}8$ (\textbf{P(i)}, right-aligned $21$-bit inset) keeps bit~$16$ inside \texttt{[19:16]} with positive sign. (3)~\texttt{normalize.v} ignores \texttt{lshift\_valid=0} when \texttt{man==0} (\textbf{D2}), emitting $-4.0/+4.23$ or $\pm\infty{\to}\mathrm{NaN}$; dashed navy \textbf{P(iii--iv)} clamps \texttt{man==0} to \texttt{16'h0}.}
  \label{fig:axcore_cascade}
\end{figure*}
Unshort-circuited differential execution of $19{,}338$ Stage-1 RV64+RoCC programs across three Chipyard configurations (SmallBoom, LargeBoom, and RocketChip) isolated $617$ return-code divergences ($0$ false CPU bug reports across $76$ negative controls). Triage deduplicated $586$ cases into six contract families plus $6$ unreported output-stationary/weight-stationary (OS/WS) transpose aborts ($25$ cases reproduce upstream issue~\#199). First, $53$ cases form silent golden-oracle escapes (Families~E--F), where \texttt{spike-gemmini} returns \texttt{PASSED} on unguarded zero-column ViT/LLM tail loads or unfenced WS RELU6 switches while Chisel RTL legitimately asserts at cycles~$44$ and~$46$; these expose missing precondition checks in \texttt{spike-gemmini} rather than RTL bugs. Second, Families~A and~D plus $6$ OS/WS cases represent over-strict reference aborts, where RTL checks the single transposer lazily or masks shift $s=32$ modulo $32$. Third, Families~B--C expose unchecked \texttt{LocalAddr} wrapping, which raises host out-of-bounds (OOB) errors in Spike yet wraps modulo 14-bit scratchpad (SPAD) in RTL, aliasing Key tile~$3$ onto tile~$0$ ($16/16$ attention flips at $139{,}397$ cycles) or redirecting negative offsets into Accumulator row \texttt{0x3f0}.

\subsection{AxCore RTL Defects, Patch, and Real AI Workload Attribution}
\label{sec:axcore_defects}

Under AxCore's design contract (MICRO~2025~\cite{axcore}, commit \texttt{b54c581}), intended approximation is bounded strictly to the approximate FP-integer multipliers (\texttt{AxMultS1}/\texttt{S2})~\cite{approxsurvey}. Our independent $82$-SLOC exact rational block-FP oracle isolates genuine RTL defects downstream of the multiplier that violate basic finite-arithmetic and zero-preservation invariants, led by coupled bugs \texttt{D1} (CWE-190 modular partial-sum wraparound) and \texttt{D2} (CWE-191 zero-mantissa exponent underflow~\cite{cwe}) in \axadder{} and \texttt{normalize.v} (while properly excluding out-of-envelope IEEE \texttt{NaN}/\texttt{Inf} case \texttt{AX-D8}). Both bugs evade the shipped FP32 CUDA simulator (\texttt{AxCore\_fp16.cu:54,119,248}), confirming that an implementation-matched simulator cannot certify arithmetic intent. Our RTL engine matches Verilator~$5.054$ and VCS bit-for-bit across $22{,}796$ vectors ($0$ mismatches).

\textbf{Bit-Level Cascade and Four-Part Hardware Patch.} Fig.~\ref{fig:axcore_cascade} traces how two coupled arithmetic bugs cascade across three pipeline stages of unmodified AxCore~\cite{axcore} commit~\texttt{b54c581}. In Stage~(1) (\axadder{}:75--85), \texttt{2's Pack} hardwires an implicit~$1$ at bit~$12$ even for zero inputs ($+2^{12}$ raw counts/lane, corresponding to $+2^{-15}$ exponent-scaled in \texttt{AX-D10}). In Stage~(2), upstream SpinalHDL hardcodes \texttt{INT=4} (17-bit signed \texttt{man\_sum[16:0]}, max $+65{,}535$) regardless of systolic depth $K{=}64$ (which requires $\lceil\log_2(2K)\rceil{=}7$ integer bits); consequently, $16$ minimal non-negative lanes sum to $16{\times}2^{12}{=}2^{16}{=}\texttt{17'h10000}$, flipping sign bit~\texttt{[16]} while zeroing \texttt{man[15:0]} (\textbf{D1}, hatched cell in the right-aligned bit-slice inset) and trapping coherent accumulations inside $[-16,+16]$ (three wraps across $K=4$--$64$, outputting $+2.69$ instead of $+98.69$). In Stage~(3) (\texttt{normalize.v}:42--112), \texttt{l\_prio\_enc} ignores \texttt{lshift\_valid=0} when \texttt{man==0} and subtracts $15 \pmod{32}$ from \texttt{exp} (\textbf{D2}), turning $60/64$ zero-mantissa encodings into spurious powers of two and $2/64$ (\texttt{exp=14}) into $\pm\infty{\to}\mathrm{NaN}$---even an all-zero $64{\times}\texttt{0x0000}$ input cascades D1 into D2, emitting $-4.0$ at rows~$4,12$ and $+4.2266$ after \texttt{AxDequant} at row~$16$.

As shown in steel-navy across Fig.~\ref{fig:axcore_cascade}, our four-part inline Verilog patch breaks the cascade at every stage: \textbf{P(i)} widens \texttt{INT} $4{\to}8$ (21-bit \texttt{man\_sum}, 26-bit \texttt{PSum}), keeping bit~$16$ cleanly inside \texttt{[19:16]=0001} with sign bit~\texttt{[20]=0}; \textbf{P(ii)} gates zero lanes via \texttt{src\_nz = |\{exp,man\}} (fixing \texttt{AX-D10}); and \textbf{P(iii--iv)} inserts a \texttt{man==0} MUX clamping outputs to \texttt{16'h0} inside \texttt{normalize.v} and \texttt{AxDequant}. Yosys~$0.60$ \texttt{sat -verify} certifies combinational D1 non-wrap ($0.59\,\mathrm{s}$; shipped refuted $0.50\,\mathrm{s}$), $2$-step BMC all-zero preservation ($0.66\,\mathrm{s}$), and $64$-encoding zero-mantissa safety ($0.04\,\mathrm{s}$). Pre-PnR $45\,\mathrm{nm}$ \nangate{} synthesis (Yosys~$0.60$, OpenSTA~$2.7.0$) incurs $+8.43\%$ cell area, $+1.32\%$ minimum period, and $+7.37\%$ power overhead on full $64{\times}64$ W4 ($+11.06\%$ area, $+6.10\%$ period, $+9.23\%$ power on W8).

\begin{figure}[t]
  \centering
  \includegraphics[width=0.82\columnwidth]{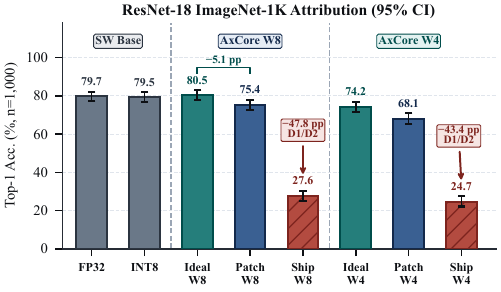}
  \caption{Expanded ResNet-18 attribution across all $1{,}000$ ImageNet-1K classes ($n=1{,}000$ images, $28.91\mathrm{B}$ RTL blocks, Wilson $95\%$ CIs). Intended approximate multiplication costs $-5.1\,$pp ($75.4\%$), whereas shipped RTL defects D1/D2 inflict a $-47.8\,$pp collapse ($27.6\%$, McNemar $p=6.8\times 10^{-126}$; $-43.4\,$pp on W4).}
  \label{fig:resnet1000}
\end{figure}

\begin{figure}[t]
  \centering
  \includegraphics[width=0.82\columnwidth]{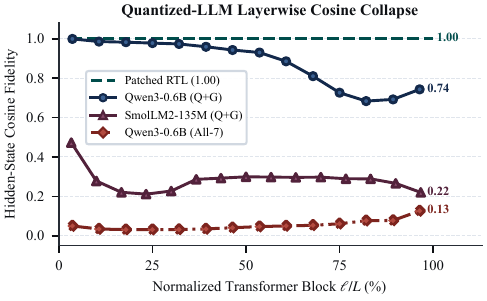}
  \caption{Layerwise hidden-state cosine fidelity on W4+QuaRot Qwen3-0.6B and SmolLM2-135M (WikiText-2, $16.4\mathrm{k}$ tokens). Patched RTL preserves $1.00$ fidelity ($0\%$ D1/D2), whereas shipped RTL collapses hidden states to $0.13$--$0.22$ and drives $100\%$ true-IEEE NaN KV-cache failure.}
  \label{fig:ai_impact}
\end{figure}

\textbf{Expanded ImageNet-1K ($n=1{,}000$) and Quantized-LLM Attribution.} Whereas zero-mean Gaussian unit tests cancel partial sums ($<0.09\%$ D1), RTL-matched replays on real pretrained weights cleanly separate quantization, intended approximation, and RTL defect impact (Figs.~\ref{fig:resnet1000}--\ref{fig:ai_impact}). On ResNet-18~\cite{resnet18} across all $1{,}000$ ImageNet-1K classes ($n=1{,}000$ images, $28.91\mathrm{B}$ RTL blocks), software FP32 ($79.7\%$, Wilson $95\%$ CI $[77.1, 82.1]$), PTQ INT8 ($79.5\%$), and Ideal W8 ($80.5\%$) establish the reference floor. Intended approximate multiplication (Patched RTL W8, $0.00\%$ D1/D2) lowers Top-1 accuracy by $-5.1\,$pp to $75.4\%$ (McNemar $p=2.03\times 10^{-7}$), whereas shipped D1/D2 ($0.94\%$ D1 / $0.0028\%$ D2 block rate) triggers a $-47.8\,$pp collapse to $27.6\%$ Top-1 ($49.2\%$ Top-5, $p=6.78\times 10^{-126}$); on W4, Ideal ($74.2\%$) $\to$ Patched ($68.1\%$, $-6.1\,$pp approx.) $\to$ Shipped drops $-43.4\,$pp to $24.7\%$ ($p=1.30\times 10^{-106}$).

On WikiText-2~\cite{wikitext} ($16{,}384$ tokens, bit-exact across $1{,}900$ Verilator vectors), shipped RTL drives up to $64.1\%$ D1 / $38.4\%$ D2 in Softmax$\cdot V$ attention ($305.7\mathrm{M}$ blocks) and $98.44\%$ D2 ($70.9\%$ NaN) on QuaRot $H_{64}$~\cite{quarot} sink/pad blocks; even a $0.003\%$ D2 rate injects $\pm\infty \to \mathrm{NaN}$ into KV caches, collapsing hidden-state cosine fidelity to $0.13$--$0.22$ (Fig.~\ref{fig:ai_impact}) and inflicting $100\%$ true-IEEE NaN perplexity failure on both Qwen3-0.6B~\cite{qwen3} and SmolLM2-135M~\cite{smollm2} ($32.59$/$18{,}505$ clamped 2-op PPL; $78.2\mathrm{k}$--$3.95\times 10^6$ clamped All-7 PPL). Our patched RTL eliminates $100\%$ of D1/D2 wraps and NaNs ($0.00\%$ rate), restoring 2-op W4+QuaRot perplexity to $26.08$ on Qwen3-0.6B (vs.\ FP32 $23.60$) and $24.52$ on SmolLM2-135M (vs.\ FP32 $19.31$), while QuaRot cuts All-7 patched PPL by $1.81\times$--$23.9\times$.

\begin{table*}[t]
\centering
\caption{Severity-ordered ledger of all $20$ newly discovered, deduplicated root causes across five DUTs ($11$ Category~N1, $3$ N2, $6$ N3; distinguishing autonomous discovery from TraceSMT/SAT targeted solving and excluding known/upstream-tracked historical recalls).}
\label{tab:forensic_audit}
\footnotesize
\setlength{\tabcolsep}{2.2pt}
\renewcommand{\arraystretch}{1.01}
\begin{tabularx}{\textwidth}{@{}l l l >{\raggedright\arraybackslash}X@{}}
\toprule
\textbf{ID} & \textbf{DUT (Commit/Cfg.)} & \textbf{Exact Source Site (File:Line)} & \textbf{Root-Cause Mechanism, Spec Basis / Impact \& Discovery Mode} \\
\midrule
\multicolumn{4}{@{}l@{}}{\textit{Category N1: Newly Discovered Functional RTL Bugs ($11$ total, severity-ordered; responsibly disclosed upstream)}} \\
CPU-BUG-05 & Rocket (\tcode{ece7b9ad}) & \tcode{CSR.scala}:473,533,1209,1284,1367 & 40b \tcode{sepc}/\tcode{mepc} violates Priv Spec \S3.1.14/\S4.1.7 (Issue \#3853; Veril.+VCS) \\
D1 & AxCore (\tcode{b54c581}) & \tcode{adder\_fpany\_no\_norm\_v2.v}:6,90--98 & \tcode{INT=4} (17b \tcode{man\_sum}) wraps at $2^{16}$; ResNet-18 $-47.8\,$pp [Patched, SAT] \\
D2 & AxCore (\tcode{b54c581}) & \tcode{normalize.v}:57--63,103,110--112 & Ignores \tcode{lshift\_valid} (\tcode{man==0}, $\mathrm{exp}{-}15\!\bmod\!32$); \tcode{exp=14}$\to\!\pm\infty$ [Pat.] \\
CVA6-D1 & CVA6 (\tcode{covO0} cfg.) & \tcode{fpnew\_divsqrt\_multi},\tcode{div\_sqrt\_top} & Inverted \tcode{C\_RM\_PLUS/MINUSINF} swaps RDN/RUP; RMM$\to$RTZ ($5/5$) \\
AX-D10 & AxCore (\tcode{b54c581}) & \tcode{adder\_fpany\_no\_norm\_v2.v}:75--76 & Implicit $1$ on zero lanes (\tcode{0x0000}): $+2^{-15}$/lane ($8.88\times$) [Patched] \\
AX-D7 & AxCore (\tcode{b54c581}) & \tcode{normalize.v}:31,34 & 2's-comp.\ abs carry-drop on \tcode{INT\_MIN} ($-65{,}536\to -0.0$/$-\infty$; SAT) \\
AX-D4\_PE & AxCore (\tcode{b54c581}) & \tcode{MPWq4/W8\_PE}:60--74,\tcode{GuardAW}:19--25 & Unclamped 16b \tcode{AxMultS1} borrow/carry flips sign (dead \tcode{SignTC}) \\
AX-D4\_Dq & AxCore (\tcode{b54c581}) & \tcode{AxDequant.scala}:23--27 & Unclamped 16b \tcode{AxMultS2} lacks $S=0$/underflow clamp ($-36\mathrm{k}$/$-\mathrm{NaN}$) \\
AX-D5 & AxCore (\tcode{b54c581}) & \tcode{normalize.v}:100--104 & 5-bit \tcode{exp} overflow wraps mod~$32$ ($4\times\mathrm{exp}=30\to\tcode{0x0000}$) \\
AX-D6 & AxCore (\tcode{b54c581}) & \tcode{normalize.v}:61--63,103 & Partial-cancel \tcode{exp} underflow wraps mod~$32$ ($\mathrm{exp}=8\to +32{,}768$; $5/5$) \\
AX-D3 & AxCore (\tcode{b54c581}) & \tcode{PreAdd.scala}:60 & \tcode{A\_Valid:=A\_FP.orR} checks bit $[15]$; $-0.0$ (\tcode{0x8000}) bypasses \tcode{GuardAW} \\
\midrule
\multicolumn{4}{@{}l@{}}{\textit{Category N2: RTL FSM Robustness, Systematic Arithmetic Bias, and HW--SW Contract Mismatches ($3$ total, severity-ordered)}} \\
GEM-D2 & Gemmini (\tcode{b82d49cc}) & \tcode{LoopMatmul}:795--807,\tcode{Util}:38--45 & No zero guard: $I/J=0$ underflows \tcode{floorAdd} to \tcode{16'hFFFF} hang \\
AX-D9 & AxCore (\tcode{b54c581}) & \tcode{adder\_fpany\_no\_norm\_v2.v}:84--86 & Two's-comp.\ ASR sticky $-1$ bias ($3.93\times$ Gauss / $6.70\times$ outlier) \\
GEM-D1 & Gemmini (\tcode{b82d49cc}) & \tcode{Configs}:27,\tcode{PE}:83,88,\tcode{Arith}:109--113 & 20b WS \tcode{spatialArrayOutputType} clips 32b OS \tcode{acc\_t} ($5/5$ vs.\ $0/5$) \\
\midrule
\multicolumn{4}{@{}l@{}}{\textit{Category N3: Reference-Simulator (\tcode{spike-gemmini} \tcode{34741e0}/\tcode{051d820f}) and SW-Header (\tcode{gemmini-rocc-tests}) Bugs ($6$ total; RTL correct)}} \\
GEM-D8 & \tcode{spike-gemmini} & \tcode{gemmini.cc}:886 (:1042 HEAD) & \tcode{bool activation=(rs2>>3)\&3} narrows \tcode{RELU6}$\to$\tcode{RELU} ($10.6\times$, \#10) \\
GEM-SW-D7 & \tcode{gemmini-rocc} & \tcode{include/gemmini.h}:373--411 & \tcode{sp\_tiled\_matmul\_os} ignores \tcode{low\_D} (fixed \tcode{acc\_t*}); $5/5$ vs.\ $0/5$ pw. \\
GEM-D5 & \tcode{spike-gemmini} & \tcode{gemmini\_params.h}:52--72,\tcode{.cc}:1525 & \tcode{ROUND\_NEAR\_EVEN} casts float$\to$long long pre-clamp (UB$\to -128$) \\
GEM-D6 & \tcode{spike-gemmini} & \tcode{gemmini.cc}:1507--1509 & 32b signed $6\ll\tcode{relu6\_shift}$ overflows at shift $29$--$31$ (RTL 35b) \\
GEM-D4 & \tcode{spike-gemmini} & \tcode{gemmini.cc}:399--408 & \tcode{config\_mvin} ignores \tcode{rs1[15:8]} \tcode{pixel\_repeats} (\tcode{conv\_dw}; v0.7+) \\
GEM-D3 & \tcode{spike-gemmini} & \tcode{gemmini.cc}:571--580 & Dead shifted var: computes \tcode{sys\_shift} (L572), writes raw (L577) \\
\bottomrule
\end{tabularx}
\end{table*}

\subsection{Specification Proofs and Unified 20-Root-Cause Ledger}
\label{sec:forensic_ledger}

\textbf{RISC-V Specification Proof (\texttt{CPU-BUG-05}) and Live Upstream 0-Day \texttt{GEM-D8}.} Rather than relying solely on reference--RTL divergence, we audited every newly isolated failure against architectural standards and live upstream repositories. First, on RocketChip and SonicBOOM, \texttt{CPU-BUG-05} (tracked upstream as \texttt{chipsalliance/rocket-chip} Issue~\#3853) violates RISC-V Privileged Spec \S3.1.14 (\texttt{mepc}) and \S4.1.7 (\texttt{sepc}): while WARL permits mapping an invalid non-canonical VA into another invalid VA (preserving \texttt{mcause=12}), it forbids silently converting an invalid non-canonical VA into a \emph{valid canonical VA}. Because \texttt{CSR.scala} stores 40-bit \texttt{reg\_sepc} without invoking \texttt{encodeVirtualAddress} (which \texttt{RocketCore.scala:1035--1040} applies on \texttt{ld}/\texttt{sd}/\texttt{jalr}), writing non-canonical \texttt{0x12345600\_800010ac} truncates bits $[63{:}40]$ into valid \texttt{0x00000000\_800010ac} and executes an S-mode payload (\texttt{mcause=9}, \texttt{tohost=0x33}, hijacked \texttt{a7=0xdeadbeefcafebabe}) on both Verilator and VCS-MX, whereas Spike and canonical/bit-39 controls fault cleanly with \texttt{mcause=12}.

Second, on Gemmini, Stage-2 \texttt{LoopConv} synthesis autonomously exposes \textbf{\texttt{GEM-D8}} in $5/5$ seeds---a live unfixed 0-day bug verified in upstream \texttt{ucb-bar/libgemmini} master HEAD (\texttt{6652649b}, \texttt{gemmini.cc:1042}; also \texttt{ucb-bar/esp-isa-sim} \texttt{051d820f}, \texttt{gemmini/gemmini.cc:1002}; disclosed as \texttt{ucb-bar/libgemmini} Issue~\#10): declaring \texttt{const bool activation = (rs2 >> 3) \& 3;} (CWE-681 boolean narrowing, compiled to x86 \texttt{setne \%dl} in \texttt{libgemmini.so}) silently truncates 2-bit \texttt{RELU6=2} to \texttt{true} (\texttt{1=RELU}) whenever hardware pooling (\texttt{!no\_pool}) re-issues \texttt{CONFIG\_ST}, dropping the $6 \ll \texttt{relu6\_shift}$ upper clamp ($127$ vs.\ $12$, a $10.58\times$ silent corruption fixed by changing \texttt{bool} to \texttt{uint8\_t}).

\textbf{Unified 20-Root-Cause Ledger (Table~\ref{tab:forensic_audit}).} Auditing every witness under unmodified upstream commits, dual simulators, Yosys SAT, and architectural specifications separates three discovery modes cleanly: \emph{Autonomous Discovery}, \emph{Known/Upstream-Tracked Recall} (excluding $6$ historical CPU benchmarks---\texttt{CPU-BUG-01} [CVE-2023-34882], \texttt{CPU-BUG-04} [CVE-2023-40065], \texttt{CPU-BUG-06}/\texttt{07}, and SonicBOOM \texttt{CPU-BUG-02}/\texttt{03}---as well as any upstream-fixed or already-tracked CVA6 errata), and \emph{Targeted Reachability Solving} (\texttt{AX-D7} via TraceSMT/SAT). Across all five designs, differential triage yields $20$ newly discovered, deduplicated root causes ($11$ functional RTL bugs, $3$ robustness/bias/contract mismatches, and $6$ reference-simulator/header bugs; excluding out-of-envelope \texttt{AX-D8} and legal unfenced-TLB events).

\section{Discussion}
\label{sec:discussion}

\subsection{Practical Takeaways and Deployment Considerations}
Dual-simulator evaluation across five RTL targets yields three engineering takeaways. First, bidirectional local--cloud collaboration overcomes both local cold-reset saturation and one-way LLM retry waste across processors and heterogeneous AI accelerators alike, winning $5/5$ seeds over FLAME on every structural, functional, and bug-finding target (Table~\ref{tab:e2e_eval}). Second, matching solver complexity to hardware depth maximizes efficiency: two-tier TraceSMT resolves shallow $1$-DFF datapaths locally in $<3.5\,\mathrm{s}$ at $\$0$ via Yosys SAT, while frozen LLM reasoning with backward Verilog slicing, VPI inversion, and compositional deep-state stacking drives closure on $100\mathrm{k}$-SLOC sequential cores and coupled accelerator schedules. Third, oracle fidelity is co-equal with stimulus generation: fast ISA models miss $53$ Gemmini assertions and harbor live narrowing bugs (\texttt{GEM-D8}), while implementation-matched FP32 simulators mask AxCore wraparounds that collapse pretrained vision and LLM accuracy.

Two practices support industrial adoption: air-gapped private-VPC deployment keeps proprietary RTL and waveforms inside the corporate perimeter, and freezing Stage-2 weights eliminates online RL drift and target-DUT fine-tuning leakage---substantiated by zero-shot CVA6 transfer and autonomous discovery of $20$ previously unreported root causes on unmodified RTL.

\subsection{Reproducibility Artifact and Responsible Disclosure}
Our reproduction package archives all pinned commits (RocketChip \texttt{ece7b9ad}, SonicBOOM \texttt{ba03175e}, Gemmini \texttt{b82d49cc}, AxCore \texttt{b54c581}, CVA6 \texttt{CVA6Config.covO0}, \texttt{libgemmini} \texttt{6652649b}, \texttt{spike-gemmini} \texttt{34741e0}/\texttt{051d820f}), tool versions (Verilator~$4.038$/$5.054$, VCS-MX~O-2018.09-SP2, Yosys~$0.60$, OpenSTA~$2.7.0$), per-seed trajectories, complete LLM call transcripts with token/USD ledgers and SHA-256 digests, SAT scripts, and $45\,\mathrm{nm}$ patch artifacts. SAT-pruned denominators ($\mathcal{P}\setminus\mathcal{P}_{\mathrm{UNSAT}}$) report sound non-refuted upper bounds, differential checking pairs structural closure with functional correctness, and responsible disclosure reports for newly discovered upstream defects (\texttt{CPU-BUG-05} [\texttt{rocket-chip}~\#3853], \texttt{AxCore D1}--\texttt{D7}/\texttt{D10}, \texttt{Gemmini D1}--\texttt{D7}, and \texttt{GEM-D8} [\texttt{libgemmini}~\#10]) have been submitted to maintainers.

\section{Conclusion}
\label{sec:conclusion}

We presented VeriSegue, a closed-loop local--cloud verification framework combining compact on-premise RL with execution-guided reasoning from frozen LLMs. Triggering handoff via toggle-immune control-flow velocity ($\gamma_{\mathrm{cf}}$) and returning deep architectural contexts to local rollouts at zero extra cloud cost overcomes both local cold-reset stagnation and one-way LLM restarting, while Stage~2 pairs admission-decoupled RTL CF-UCB, \texttt{SynLib} constructor synthesis, two-tier TraceSMT, and $K=5$ boundary batching under a strict three-oracle gate. Across five designs, eight LLMs, and two simulators (Verilator and VCS), VeriSegue achieves unanimous five-seed full-horizon gains over FLAME while exposing $20$ newly discovered root causes ($11$ RTL bugs plus live upstream 0-day \texttt{GEM-D8}, \texttt{libgemmini}~\#10), $53$ silent Gemmini escapes, and a formally checked $45\,\mathrm{nm}$ AxCore patch. More broadly, substantiated by our empirical findings, we view closed-loop local--cloud collaboration as a promising direction across chip design and verification---including RTL exploration, assertion synthesis, multi-IP coherency, and debug triage---keeping proprietary RTL and waveforms local for R\&D privacy while uniting fast local exploration with bounded cloud reasoning for faster convergence and stronger effectiveness.

\section*{Acknowledgments}
This work was supported by the National Key R\&D Program of China (2022YFB4500402). The authors would like to thank the anonymous reviewers for their valuable feedback.

\bibliographystyle{IEEEtran}
\bibliography{references_ieee}

\end{document}